\documentclass[12pt]{article}
\usepackage[utf8]{inputenc}
\usepackage{graphicx}

\title{Credible fusion of evidence in distributed system subject to cyberattacks\footnote{This work has been submitted to the IEEE for possible publication. Copyright may be transferred without notice, after which this version may no longer be accessible.}}
\author{Chaoxiong Ma \thanks{chaoxiongma@mail.nwpu.edu.cn} \and Yan Liang \thanks{liangyan@nwpu.edu.cn}}

\usepackage{subcaption}
\usepackage{multirow}
\usepackage{array}
\usepackage{algorithm}
\usepackage{algorithmic}
\usepackage{amsthm}
\usepackage{amsmath}
\usepackage{amsfonts}
\usepackage{amssymb}
\usepackage{booktabs}
\usepackage{caption}
\usepackage{enumitem} 
\usepackage{float}
\usepackage [colorlinks, linkcolor=blue, anchorcolor=blue, citecolor=blue, urlcolor=blue] {hyperref}
\usepackage{longtable}
\usepackage[english]{babel}
\usepackage[figuresright]{rotating}

\newtheorem{definition}{Definition}[section]
\newtheorem{pf}{Proof}
\newtheorem{thm}{Theorem}
\newtheorem{assumption}{Assumption}

\usepackage[dvipsnames]{xcolor}
\usepackage{bm}
\hypersetup{colorlinks,urlcolor=blue}
\usepackage{svg}
\usepackage{adjustbox}

\usepackage[numbers,sort&compress]{natbib}
\newcolumntype{R}[2]{%
    >{\adjustbox{angle=#1,lap=\width-(#2)}\bgroup}%
    l%
    <{\egroup}%
}

\date{} 

\begin{document}

\def\spacingset#1{\renewcommand{\baselinestretch}%
{#1}\small\normalsize} \spacingset{1}

\maketitle
Given that distributed systems face adversarial behaviors such as eavesdropping and cyberattacks, how to ensure the evidence fusion result is credible becomes a must-be-addressed topic. Different from traditional research that assumes nodes are cooperative, we focus on three requirements for evidence fusion, i.e., preserving evidence's privacy, identifying attackers and excluding their evidence, and dissipating high-conflicting among evidence caused by random noise and interference. To this end, this paper proposes an algorithm for credible evidence fusion against cyberattacks. Firstly, the fusion strategy is constructed based on conditionalized credibility to avoid counterintuitive fusion results caused by high-conflicting. Under this strategy, distributed evidence fusion is transformed into the average consensus problem for the weighted average value by conditional credibility of multi-source evidence (WAVCCME), which implies a more concise consensus process and lower computational complexity than existing algorithms. Secondly, a state decomposition and reconstruction strategy with weight encryption is designed, and its effectiveness for privacy-preserving under directed graphs is guaranteed: decomposing states into different random sub-states for different neighbors to defend against internal eavesdroppers, and encrypting the sub-states' weight in the reconstruction to guard against out-of-system eavesdroppers. Finally, the identities and types of attackers are identified by inter-neighbor broadcasting and comparison of nodes' states, and the proposed update rule with state corrections is used to achieve the consensus of the WAVCCME. The states of normal nodes are shown to converge to their WAVCCME, while the attacker's evidence is excluded from the fusion, as verified by the simulation on a distributed unmanned reconnaissance swarm.


\section{Introduction}\label{sec:ECFAC_Introduction}
With the rapid development of technologies such as sensors, computer science, and communication, distributed information fusion technology has been drawing more and more attention in both academics and industry. Its typical applications include vehicle cooperative sensing \cite{zoghby2014evidential,cai2023consensusbased}, maneuvering target tracking \cite{li2024distributedjoint,hu2024resource}, smart grid scheduling \cite{xu2023aprivacysmatrgrid,li2023distributedsmartgrid} and so on. Different from the traditional centralized system that pools and fuses multi-source information at a specific fusion center, the distributed system exchanges information only between neighboring intelligent nodes that integrate sensing, computing, and communication functions together, and reaches a fusion consensus using well-designed communication and data processing strategies.

It is well known that the representation and processing of uncertain data has always been the key to information fusion. As a decision-level information fusion theory, Dempster-Shafer (DS) theory  \cite{shafer1976amathematical} is already utilized in the fields of pattern recognition \cite{geng2021earc,jiao2015belief}, multi-objective association \cite{laghmara2020heterogeneous}, multi-criteria decision making \cite{zhu2021fuzzy}, and risk analysis \cite{pan2020multi}, and so on over the past three decades, and demonstrates the advantages of convenience, flexibility, and good interpretability. This motivates this paper to focus on the study of evidence fusion algorithms. The theory transforms uncertain data into evidence about the topic of interest and fuses them with Dempster's combination rule. However, Dempster's combination rule may yield a counterintuitive result while fusing high-conflict evidence caused by source failures or environmental disturbances. Related studies propose two types of improvements: alternative combination rules \cite{smets1990thecombination,dubois1988representation,yager1987onthedempster} and evidence-preprocessing-based approaches \cite{murphy2000combining,xiao2019multi,liu2011combination}. The latter ensures that the fusion results are ``credible'' by reducing conflicts based on the credibility obtained by comparing pieces of evidence and shows greater generalizability than the former in numerous engineering practices.

Recently, researchers began to focus on distributed evidence fusion. Initially, Dempster's combinatorial rule is directly applied to fusing evidence from distributed nodes because it satisfies the associative law and the commutative law. This approach faces the cyclic propagation of intermediate fusion results through the network, i.e., there is a risk of the same evidence being accumulated multiple times. After analyzing the Dempster combination rule under the representation of the commonality function deeply, an evidence fusion algorithm based on linear average consensus is proposed, which effectively solves the problem of circular data propagation \cite{kanjanatarakul2017distributed}. Some studies pre-processed evidence with credibility assessed with a priori information such as sensor performance \cite{cheng2020brb}, environmental disturbances \cite{cheng2021health}, historical failure data \cite{pietro2015situational}, fixation values, etc., which displays a credibility-evidence mismatch in most cases. The multi-stage distributed evidence fusion algorithm based on RANSAC is proposed for solving the counterintuitive knowledge problem under the distributed fusion paradigm \cite{denoeux2021distributed}. It first removes the anomalous evidence according to the conflict measurement between the to-be-fused evidence and the reference bases that are obtained by fusing randomly selected evidence, while the unremoved evidence is fused to generate the final result. By replacing the RANSAC with the connectivity-based outlier factor (COF) strategy, the network-wide random selection of evidence in the anomalous evidence rejection process is optimized, which significantly reduces the computational complexity \cite{zhao2023information}.

The two studies of detecting anomalous evidence mentioned above do not dissipate the conflict among randomly selected evidence, which may generate counterintuitive fusion results as the reference bases deviate from the ground truth and therefore do not ensure the fusion result is credible. In addition, all available studies assume that all network nodes cooperate with each other. However, distributed systems in real-world engineering environments often face cyberattacks such as information eavesdropping \cite{chen2023privacy}, denial-of-service (DoS) attacks \cite{wang2024important}, and deception attacks \cite{zhao2024dosandsteal}, which pose new challenges about the credibility of fusion. On the one hand, eavesdroppers analyze the nodes' preferences with directly received or indirectly inferred raw evidence, forcing the nodes to limit the scope and content of information sharing, which makes the sharing of information in the fusion process inadequate. On the other hand, the DoS attack prevents normal nodes from accessing the data packets of the attacked nodes by consuming critical resources such as computility and communication bandwidth; the deception attack prevents the nodes from reaching consensus by hijacking and tampering with the data packets. These attacks force existing algorithms based on anomalous evidence detection to decompose the fusion process into multiple serial consensus phases, which limits the scalability and flexibility of the algorithms and leads to the accumulation of attack risks. The attacker may perform a denial of service or inject false data at any consensus stage.

This paper studies the credible fusion of evidence in distributed systems subject to cyberattacks, including both eavesdropping and attack behaviors. Eavesdroppers are defined here as malicious entities that attempt to illegally collect or infer raw evidence of nodes, either from the outside of the system or as system nodes. Attack behaviors involve DoS attacks and deception attacks. The system nodes that are directly acted upon by these attacks are referred to as attackers, while the others are labeled as normal nodes. According to the discussion above, it can be seen that realizing credible evidence fusion needs to ensure three levels of credibility at the same time: assessing the credibility of normal nodes' evidence is the guarantee for avoiding counterintuitive errors; ensuring the credibility of information transmission is the basis of nodes' collaboration; and detecting the credibility of nodes is the key to accurately reaching the fusion consensus. To this end, this research aims to develop an algorithm for credible evidence fusion against cyberattacks (CEFAC). The contributions of this paper are:
\begin{enumerate}[label=(\arabic*)]
	\item{To avoid the fusion of normal node evidence from producing counterintuitive results, a distributed evidence fusion architecture is designed based on the conditionalized evidence credibility, under which the key to fusing multi-source evidence is proven to be obtaining the weighted average value by conditional credibility of multi-source evidence (WAVCCME). The distributed evidence fusion is thus decomposed into two steps: all nodes collaboratively calculate the WAVCCME, and nodes independently get the fusion result. It is proven that obtaining WAVCCME is equivalent to calculating the sum of the Kronecker product for evidence and event-support vector (KPEEV) and the sum of the event-support vector for evidence (EVE). This greatly simplifies the process of distributed credible evidence fusion while facilitating the design of cyberattacks defense strategy.}
	\item{As sharing KPEEV and EVE with neighbors exposes the node's raw evidence, this paper develops a state decomposition and reconstruction strategy for weighted encryption to protect the privacy of the evidence. This strategy decomposes the initial state of the node containing KPEEV and EVE into multiple random sub-states that are sent to different neighbors, respectively, and the reconstructed state is obtained by aggregating the random sub-states collected from neighboring nodes. The distributed computation of WAVCCME can take the reconstructed state as the initial state because the state sum of all nodes before and after the reconstruction is kept the same. The randomized decomposition of initial states solves the problem that traditional state decomposition strategies cannot be applied to directed graphs, and the weight encryption of random sub-states effectively protects the privacy of the raw evidence.}
	\item{To exclude the evidence of attackers from the fusion, an attacker's identity/type identification and compensation strategy is proposed. The recognition of attacker identity and type is realized by inter-neighbor broadcasting and comparison of node states, on which normal nodes can perform tailored state corrections to achieve the consensus of WAVCCME.}
\end{enumerate}

The paper's structure is outlined as follows. In Section \ref{sec:CEFAC_ProblemFormulation}, the problem studied in this paper is explored. In Section \ref{sec:CEFAC_AlgorithmDesign}, the design details of CEFAC are presented. In Section \ref{sec:CEFAC_Simulation}, the CEFAC is compared with existing algorithms, and the robustness of the proposed method to network attacks is verified by the case of a distributed unmanned reconnaissance cluster. Section \ref{sec: conclusion} concludes the study.

\section{Problem formulation}
\label{sec:CEFAC_ProblemFormulation}
Considering the collaborative decision-making requirements for the cluster composed of $N$ unmanned intelligent agents in a complex electromagnetic confrontation environment, this paper focuses on fusing multi-source evidence resiliently and credibly in distributed systems. Noting that factors, such as accusation relationships, differences in detection and communication capabilities, and electromagnetic attacks and defenses, often destroy the bi-directionality of information flow between nodes \cite{hadjicostis2016robust}, this paper utilizes the directed graph ${\mathcal{G}}=\left({\mathcal{V}},{\mathcal{E}},{\mathcal {A}}\right)$ models the distributed system, where ${\mathcal{V}}=\{1,2,\cdots,N\}$ and ${\mathcal{E}}\subseteq{\mathcal{V}}\times{\mathcal{V}}$ are the set of nodes and the set of edges, respectively. The adjacency matrix ${\mathcal {A}}$ is the matrix representation of ${\mathcal{E}}$. If $\left(j,i\right)\in\mathcal{E}$, Node $i$ can receive information from Node $j$, and $a_{ij}=1$. Otherwise, Node $i$ cannot receive information from Node $j$ and $a_{ij}=0$. In general, $a_{ii}=0$, i.e., the graph is acyclic. All nodes that can send information directly to Node $i$ constitute the set of incoming neighbor nodes ${\mathcal{N}}_i^-=\left\{j\in{\mathcal{V}}|\left(j,i\right)\in{\mathcal{E}}\right\}$ of Node $i$. All nodes that can directly receive information from Node $i$ constitute the set of outgoing neighbor nodes ${\mathcal{N}}_i^+=\left\{j\in{\mathcal{V}}|\left(i,j\right)\in{\mathcal{E}}\right\}$ of Node $i$. The in-degree and out-degree of Node $i$ are $\left|{\mathcal{N}}_i^-\right|$ and $\left|{\mathcal{N}}_i^+\right|$, respectively. In addition, the information held by a node at a given point in time is called the state of the node.

Nodes transform measurement data for decision-making into evidence defined on the same framework of discrimination (FoD) and fuse them to deal with measurement uncertainty. The FoD is a finite set of mutually exclusive and complete events \cite{zhao2020risk}, denoted $\Omega=\{\hat{A}_1,\hat{A}_2,\cdots,\hat{A}_n\}$. As the common representation form of evidence, the mass function is a mapping from the power set of the FoD $2^\Omega\triangleq\{A|A\subseteq\Omega\}$ to the interval $[0,1]$, i.e., $m(\cdot):2^\Omega\rightarrowtail[0,1]$ \cite{shafer1976amathematical}:
\begin{equation}
	\sum\nolimits_{A \subseteq \Omega } m\left( A \right) = 1
	\label{eq:Introduction_MassFunctionDefinition}
\end{equation}
where $m(A)$ denotes the probability that the set of events $A$ is true, also called the mass value of $m(\cdot)$ for the set of events $A$ or basic belief assignment (BBA). If cardinality $|A|=1$, then $A$ is said to be a singleton class; otherwise, it is called a compound class. This paper adopts the $2^{|\Omega|}$-dimensional vector $\boldsymbol{m}=[m(\emptyset),m(A_1),m(A_2),\cdots,m(A_{2^{|\Omega|}-1})]^T$ to denote evidence. A mass function satisfying $m(\emptyset)\!>\!0$ is said to be subnormal and normalized otherwise. The normalization operation transforms the subnormal mass function $m^*$ into the normalized mass function $\boldsymbol{m}$:
\begin{equation}
	\forall A \subset \Omega, A\neq \emptyset : m(A) = \frac{m^*(A)}{1-m^*(\emptyset)}
	\label{eq:Introduction_MassFunctionNormalization}
\end{equation}

Since the empty set cannot happen in the real world, the $m(\emptyset)$ in $\boldsymbol{m}$ is omitted by default in the following content. The Dempster's combination rule is used to fuse two mutually independent pieces of evidence $\boldsymbol{m}_1$ and $\boldsymbol{m}_2$ that are defined under the same FoD $\Omega$. $\forall A \in 2^{\Omega}$:
\begin{equation}
	\left({m_1}\oplus{m_2}\right)\left(A\right) = 
	\begin{cases}
		0 &, \text{if} A=\emptyset \\
		\frac{\sum\limits_{B\cap{C=A}} {m_1\left(B\right)m_2\left(C\right)} }{1-K} &, \text{if} A \in 2^\Omega \backslash \{\emptyset\}
	\end{cases}
	\label{eq:Introduction_DempsterCombinationRule}
\end{equation}
where $\oplus$ is the Dempster fusion operator and $K$ is the degree of conflict between $m_1$ and $m_2$:
\begin{equation}
	K={\sum _{{B}\cap{C}=\emptyset}}{m_1}\left(B\right){m_2}\left(C\right)
	\label{eq:Introduction_ConfliectDegree}
\end{equation}

As described in the Introduction, three types of cyberattacks—eavesdroppers, DoS attacks, and spoofing attacks—are considered in this paper. An eavesdropper either may be a system node that can only receive information from incoming neighbors or may be an entity external to the system that can collect information transmitted on all edges at the same time. The DoS and deception attacks, although initiated by entities external to the system, directly attack the nodes rather than the edges. Therefore, system nodes that are acted upon by DoS and deception behaviors are referred to as attackers. Denote the set of all attackers as ${\mathcal{V}}_c$, the set of remaining normal nodes as ${\mathcal{V}}_n$, the set of DoS attackers as ${\mathcal{V}}_{DoS}$, and the set of deception attackers as ${\mathcal{V}}_{D}$. It is clear that there are ${\mathcal{V}}={\mathcal{V}}_c\cup{\mathcal{V}}_{n}$ and ${\mathcal{V}}_c={\mathcal{V}}_{DoS}\cup{\mathcal{V}}_{D}$. In addition, this paper also makes the following assumptions about the specific form of the attack:
\begin{assumption}
	The sparsity of the type of attacker. In the evidence consensus, a node suffers at most one type of attack. The total number of attackers is equal to the sum of the number of DoS attackers and the number of deception attackers, i.e., there is $|{\mathcal{V}}_c|=|{\mathcal{V}}_{DoS}|+|{\mathcal{V}}_{D}|$.
\end{assumption}
\begin{assumption}
	The consistency of attackers' output. Any deception attacker sends out the same message to all neighbor nodes of itself.
\end{assumption}
\begin{assumption}
	The persistence of attackers' behavior. During the fusion process, the attack type of any node remains unchanged, and the attack behavior goes on throughout the fusion process.
	\label{AttackCEF_Assumption_3}
\end{assumption}
\begin{assumption}
	The robustness of the network under attack. The directed network consisting of all nodes except attackers is strongly $p$-fraction robust:
\end{assumption}
Network robustness is usually utilized to depict the ability of distributed systems to keep their functionality and performance when suffering from attacks, failures, or anomalous inputs.
\begin{definition}[$p$-fraction reachable set]
	For a non-empty subset ${\mathcal{S}}\subseteq{\mathcal{V}}$ of graph ${\mathcal{G}}$, if $\exists{i}\in{\mathcal{S}}$ such that $\left|{\mathcal{N}}_i^-\setminus{\mathcal{S}}\right|\ge{p}\left|{\mathcal{N}}_i^-\right|>0$ with ${p}\in[0,1]$, then ${\mathcal{S}}$ is said to be $p$-fraction reachable.
\end{definition}
\begin{definition}[$p$-fraction robust graph]
	For a nonempty and nontrivial directed graph ${\mathcal{G}}$ with at least two nodes, ${\mathcal{V}}$ is said to be $p$-fraction robust if there is at least one $p$-fraction reachable set of any pair of disjoint nonempty subsets of ${\mathcal{V}}$. Correspondingly, all empty or trivial directed graphs ${\mathcal{G}}$ are $p$-fraction robust.
\end{definition}
\begin{definition}[Strongly $p$-fraction robust graph]
	A directed graph ${\mathcal{G}}$ is said to be strongly $p$-fractionally robust if any of its nonempty subsets ${\mathcal{S}}\subseteq{\mathcal{V}}$ is $p$-fraction reachable or $\exists{i}\in{\mathcal{S}}$ makes $\left|{\mathcal{V}}\setminus{\mathcal{S}}\right|\!\subseteq\!{\mathcal{N}}_i^-$.
\end{definition}

Existing attack models, such as the $f$-global attack model \cite{leblanc2013resilient} and the $f$-local attack model \cite{zhang2012robustness}, assume there is an upper bound on the number of attackers. The normal nodes affected by an attacker may be more or less, since the difference in (outgoing/incoming) degrees among nodes scales up with the network size. In other words, these models may fail to accurately portray the situation faced by normal nodes. This paper defines the attacker number model \cite{ying2023qipian}in the form of the percentage of attackers in the neighborhood:
\begin{definition}[$f$-fraction local cyberattacks model]
	The distributed system ${\mathcal{G}}$ is said to be under $f$-fraction local cyberattack if $\forall{i}\in{\mathcal{V}}_n$, $\left|{\mathcal{V}}_c\right|\leq{f}\left|{\mathcal{N}}_i^-\right|$.
\end{definition}
Three requirements for the fusion of $N$ pieces of evidence and the challenges faced by these requirements are given below under the assumptions mentioned above:
\begin{enumerate}[label=(\arabic*)]
	\item Directly applying Dempster's rule to fuse highly conflicting multi-source evidence caused by sensor interference or system bias generates counterintuitive results, so fusion algorithms must ensure ``credibility'' at the level of the measured information. However, the localized sharing of node information and the requirement of privacy protection make the evidence credibility assessment complex, inconsistent, and unusable. Therefore, objectively and accurately assessing the fusion weight of evidence becomes the primary challenge in algorithm design.
	\item Nodes are reluctant to disclose their evidence due to concerns about their data privacy. However, existing evidence credibility assessment inevitably requires neighbors to share their raw evidence with each other. How to prevent eavesdroppers from estimating the raw evidence $\boldsymbol{m}_i$ held by any Node $i\in{\mathcal{V}}$ with arbitrary accuracy, so as to ensure that the fusion is credible at the level of information transmission, has become a major challenge for algorithm design.
	\item The potential attackers would prevent normal nodes from reaching the fusion consensus: $\boldsymbol{m}_\oplus=\boldsymbol{m}^1_\oplus=\boldsymbol{m}^2_\oplus=\cdots=\boldsymbol{m}^{|{\mathcal{V}}_n|}_\oplus$, where $\boldsymbol{m}^i_\oplus$ denotes the fusion result obtained by Node $i$. Thus, the fusion algorithm must ensure the fusion participant is credible. How to improve the capabilities of error detection and fault tolerance to exclude the attacker's evidence from the fusion becomes a challenge for algorithm design.
\end{enumerate}

The three levels of credibility are interdependent and should be considered together. The prerequisite for accurately assessing the credibility of evidence is to exclude the influence of malicious nodes. Meanwhile, protecting the privacy of normal nodes helps to identify attackers.

\section{Distributed credible evidence fusion against cyberattacks}
\label{sec:CEFAC_AlgorithmDesign}
A distributed credible evidence fusion algorithm against cyberattacks is proposed in this section. It prevents attackers from interfering with the fusion results and also overcomes the high conflict paradox. First, the above evidence fusion task is modeled based on conditionalized credibility, and it is proved that fusing multi-source evidence is equivalent to obtaining WAVCCME, which leads to the conclusion that the key for distributed credible fusion lies in the consensus of KPEEV and EVE. After that, the state decomposition and reconstruction strategy with weight encryption is proposed to guarantee information transmission is credible during the consensus of KPEEV and EVE. Finally, an attacker identity/type identification and compensation strategy is designed to ensure that the nodes participating in the fusion are credible.
\subsection{Key to fusion: the WAVCCME}\label{subsec:WAVCCME_Introduce}
Traditional centralized credible evidence fusion, such as \cite{liu2023conflict}, calculates the credibility by two-by-two comparison between pieces of evidence. This credibility assessment method is hardly adapted to the requirements of sharing information among neighbors and privacy protection in distributed systems. To this end, this subsection employs conditionalized credibility as evidence fusion weights to model the multi-source evidence fusion task. Conditionalized credibility $Cerd_i\triangleq{p(c_i|{\hat{A}}_j)}$ is the likelihood of the event $c_i$, i.e., ``$\boldsymbol{m}_i$ is the most plausible evidence", occurring when ${\hat{A}}_j$ is taken as the decision value:
\begin{equation}
	Cerd_i=p\left(c_i\right)=p\left(c_i,\Omega\right)= \sum\limits_{j=1}^n{p}\left(c_i|{\hat{A}}_j\right)p\left({\hat{A}}_j\right)
	\label{eq:ICEF_CredConditionalExpand}
\end{equation}
The evidence fusion task can be described in the following iterative form:
\begin{align}
	&Cerd_i^{\left(t\right)}= \sum\limits_{j=1}^n{p}\left(c_i|\hat{A}_j\right)p^{\left(t-1\right)}\left(\hat{A}_j\right)\label{eq:ICEF_Summary_1}\\
	&\boldsymbol{m}_{avg}^{(t)}= \sum\limits_{i=1}^N{{\mathfrak{f}}(\mathfrak{N}_i)Cerd_i^{(t)}\boldsymbol{m}_i}\label{eq:ICEF_Summary_2}\\
	&\boldsymbol{m}_{\oplus}^{(t)}= \underbrace{\boldsymbol{m}_{avg}^{(t)}\oplus\boldsymbol{m}_{avg}^{(t)}\oplus\cdots\oplus\boldsymbol{m}_{avg}^{(t)}}_{|{\mathcal{V}}_n|-1}\label{eq:ICEF_Summary_3}\\
	&p^{\left(t\right)}\left(\hat{A}_j\right)= BetP_{{{\boldsymbol{m}}_\oplus}^{\left(t\right)}}\left(\hat{A}_j\right)\label{eq:ICEF_Summary_4}
\end{align}
where $(t)$ denotes the iteration step, and ${\mathfrak{f}}(\mathfrak{N}_i)$ is an indication of node normality, which is 0 when the Node $i$ is an attacker, and 1 otherwise. $BetP_{\boldsymbol{m}}(\hat{A}_j)$ denotes the transformation function of the mass function to the Pignistic probability:
\begin{equation}
	BetP_{\boldsymbol{m}}\left(\hat{A}_j\right)=\sum\limits_{\hat{A}_j \subseteq{B} \subseteq\Omega}\frac{m\left(B\right)}{\left|B\right|}
	\label{eq:Introduction_MassToBetP}
\end{equation}
Conditional credibility is computed from the support $sup_{ji}$ of $\boldsymbol{m}_{i}$ for $\boldsymbol{m}_{\hat{A}_{j}}$:
\begin{equation}
	p\left(c_i|{\hat{A}}_j\right)=\frac{sup_{ji}}{\sum\limits_{{i}\in{{\mathcal{V}}_n}}sup_{ji}}=\frac{e^{-\tau{d}_{ji}}}{\sum\limits_{{i}\in{{\mathcal{V}}_n}}e^{-\tau{d}_{ji}}}
	\label{eq:ICEF_ConditionalCredDef}
\end{equation}
where $\tau$ is the distance coefficient, and $d_{ji}$ is the evidence difference measure between the to-be-fused evidence $\boldsymbol{m}_i$ and the event evidence $\boldsymbol{m}_{\hat{A}_j}$, whose computation can be referenced to the \cite{xiao2019multi,xiao2020anew, pan2022enhanced}. Here, the event evidence $\boldsymbol{m}_{\hat{A}_j}$ indicates that ``the event $\hat{A}_j$ is the decision truth value”:
\begin{equation}
	m_{\hat{A}_j}\left(B\right) = 
	\begin{cases}
		1 &, \text{if}\ B = \hat{A}_j \\
		0 &,{\text{otherwise.}}
	\end{cases}
	\label{eq:ICEF_EventEvidence}
\end{equation}

Eqs.(\ref{eq:ICEF_Summary_1})-(\ref{eq:ICEF_Summary_4}) describe the credibility at the level of the measurement information and at the level of the source in terms of $Cred_i$ and $\mathfrak{f}(\mathfrak{N}_i)$, respectively. However, the calculation of the conditionalized credibility in Eq.(\ref{eq:ICEF_Summary_1}) relies on the sum of support of all evidence $\sum_{i=1}^{|{{\mathcal{V}}_n}|}sup_{ji}$. The identification of the attacker in Eq.(\ref{eq:ICEF_Summary_2}) and the counting of the number of normal nodes in Eq.(\ref{eq:ICEF_Summary_3}) are also coupled with each other. All of these are disadvantageous to the design of distributed fusion algorithms. Further analysis around how to implement Eqs.(\ref{eq:ICEF_Summary_1})-(\ref{eq:ICEF_Summary_4}) in a distributed manner is presented below.

Let $P(c_i|A)=[p(c_i|\hat{A}_1),p(c_i|\hat{A}_2),\cdots,p(c_i|\hat{A}_n)]\in{\mathbb{R}^{{1}\times{n}}}$ be the row vector composed of the overall conditional credibility of $\boldsymbol{m}_i$. The WAVCCME of all normal nodes is:
\begin{equation}
	\boldsymbol{m}_{avg|A}\triangleq\sum\limits_{i=1}^{|{\mathcal{V}}_n|}\boldsymbol{m}_{i}\otimes{P(c_i|A)}
	\label{eq:AttackCEF_ConditionalCredMassAvg}
\end{equation}
where $\otimes$ is the Kronecker product operator. The $\boldsymbol{m}_{avg|A}$ is a matrix of $(2^{|\Omega|}-1)$ rows and $n$ columns. $\forall{j}\in\{1,2,\cdots,n\}$, $\sum_{i=1}^{|{\mathcal{V}}_n|}{p(c_i|\hat{A}_j)}=1$. Thus any column of $\boldsymbol{m}_{avg|A}$ is a mass function: the weighted average of all to-be-fused evidence when the event $\hat{A}_i$ is decision-true and satisfies all of the properties possessed by a mass function.

\begin{thm}
	\label{theorem_1}
	If one obtains $\boldsymbol{m}_{avg|A}$, it is equivalent to obtaining the fusion result of the evidence of all normal nodes according to Eqs.(\ref{eq:ICEF_Summary_1})-(\ref{eq:ICEF_Summary_4}).
\end{thm}
\begin{pf}
	To prove Theorem \ref{theorem_1}, let's assume that the exact value of the identity indicator ${\mathfrak{f}}(\mathfrak{N}_i)$ of all nodes is known in advance. The joint Eqs.(\ref{eq:ICEF_Summary_1})-(\ref{eq:ICEF_Summary_4}) yields:
	\begin{equation}
		\begin{aligned}
			\boldsymbol{m}_\oplus \! &= \! \mathop  \oplus \limits_{j = 1}^{{|{\mathcal{V}}_n|} - 1} \boldsymbol{m}_{avg}\\
			&=\!\mathop  \oplus \limits_{j = 1}^{{|{\mathcal{V}}_n|} - 1} \left( \sum\limits_{i = 1}^{|{\mathcal{V}}_n|} \boldsymbol{m}_i \!\otimes\! P\left(c_i|A\right)^T BetP{\left(\frac{\boldsymbol{m}_\oplus}{1 - m_\oplus (\emptyset )}\right)}\right)\\
			&=\!\mathop  \oplus \limits_{j = 1}^{{|{\mathcal{V}}_n|} - 1} \left(\boldsymbol{m}_{avg|A}BetP{\left(\frac{\boldsymbol{m}_\oplus}{1 \!-\! m_\oplus (\emptyset )}\right)}\right)
		\end{aligned}
		\label{eq:AttackCEF_ICEFSummary}
	\end{equation}
	Clearly, Eq.(\ref{eq:AttackCEF_ICEFSummary}) is a nonlinear equation about $\boldsymbol{m}_\oplus$, and the only factor that determines the value of $\boldsymbol{m}_\oplus$ is $\boldsymbol{m}_{avg|A}$.
\end{pf}
According to Theorem \ref{theorem_1}, the only thing to realize distributed evidence fusion is that each node obtains the WAVCCME of all normal nodes. Therefore, distributed evidence fusion can be decomposed into two sub-tasks: the WAVCCME consensus, and the local solution of the fusion result based on the WAVCCME. The difficulty for the WAVCCME consensus is how to calculate $p(c_i|\hat{A}_j)$. According to Eq.(\ref{eq:ICEF_ConditionalCredDef}), the computation of $p(c_i|\hat{A}_j)$ relies on $\{sup_{ji}|{i}=1,2,\cdots,{|{\mathcal{V}}_n|}\}$. In other words, the node needs to obtain the evidence of all normal nodes at the same time. However, nodes are reluctant to share their raw evidence for privacy-preserving reasons, which provokes the demand to develop distributed strategies for computing the conditional credibility of evidence.

\begin{thm}
	\label{theorem_2}
	Consensus normal nodes' WAVCCME is equivalent to consensus their KPPEEV $\bar{X}$ and EVE $\bar{Y}$:
	\begin{align}
		\bar{X} \triangleq & \left[ \sum_{i = 1}^{|{\mathcal{V}}_n|}  sup_{1i}\boldsymbol{m}_i,
		\sum_{i = 1}^{|{\mathcal{V}}_n|}   sup_{2i}\boldsymbol{m}_i,\cdots,
		\sum_{i = 1}^{|{\mathcal{V}}_n|}   sup_{ni}\boldsymbol{m}_i\right]^T \label{eq:AttrackCEF_XDef}\\
		\bar{Y} \triangleq & \left[ \sum_{i = 1}^{|{\mathcal{V}}_n|}   sup_{1i},
		\sum_{i = 1}^{|{\mathcal{V}}_n|}   sup_{2i},\cdots,
		\sum_{i = 1}^{|{\mathcal{V}}_n|}   sup_{ni}\right]^T 
		\label{eq:AttrackCEF_YDef}
	\end{align}
\end{thm}
\begin{pf}
	Substituting Eq.(\ref{eq:ICEF_ConditionalCredDef}) into Eq.(\ref{eq:AttackCEF_ConditionalCredMassAvg}) yields:
	\begin{equation}
		\begin{aligned}
			\boldsymbol{m}_{avg|A} = &\sum\limits_{i = 1}^{|{\mathcal{V}}_n|}  \boldsymbol{m}_i   \otimes \left[ \frac{sup_{1i}} {\sum\limits_{i = 1}^{|{\mathcal{V}}_n|}  sup_{1k}}, \frac{sup_{2i}} {\sum\limits_{k = 1}^{|{\mathcal{V}}_n|}  sup_{2k}}, \cdots ,\frac{sup_{ni}} {\sum\limits_{k = 1}^{|{\mathcal{V}}_n|}  sup_{nk} }\right]\\
			= &\sum\limits_{j = 1}^n \frac{\sum_{i = 1}^{|{\mathcal{V}}_n|}    sup_{ji}\boldsymbol{m}_i}{\sum_{k = 1}^{|{\mathcal{V}}_n|}  sup_{jk}}\\
			= & \textbf{1}^T \text{diag}(\bar{Y})^{-1}\bar{X}
		\end{aligned}
		\notag
	\end{equation}
	where $\text{diag}(\bar{Y})$ is the diagonal matrix with $\bar{Y}$ as the main diagonal elements. Therefore, if the distributed nodes can collaboratively consensus on $\bar{X}$ and $\bar{Y}$, they are equivalent to obtaining $\boldsymbol{m}_{avg|A}$.
\end{pf}
For the consensus of $\bar{X}$ and $\bar{Y}$, the information that can be provided by any Node $i$ consists of $Y_i$: the event support vector (ESV) of $\boldsymbol{m}_i$, and $X_i$: the direct product of evidence $\boldsymbol{m}_i$ and its ESV (DPEESV):
\begin{align}
	Y_i(0)=&\left[sup_{1i},sup_{2i},\cdots,sup_{ni}\right]^T\label{eq:AttrackCEF_YiDef}\\
	X_i(0)=&Y_i(0)\otimes\boldsymbol{m}_i \label{eq:AttrackCEF_XiDef}
\end{align}
Obviously, there is:
\begin{align}
	\bar{X}=&\sum\limits_{i=1}^{|{\mathcal{V}}_n|}X_i(0) \label{eq:relationBarX_Xi}\\
	\bar{Y}=&\sum\limits_{i=1}^{|{\mathcal{V}}_n|}Y_i(0) \label{eq:relationBarY_Yi}
\end{align}
Let $x_i(t)\triangleq\{X_i(t),Y_i(t)\}$ denote the state of Node $i$ at moment $t$. The key of the distributed evidence fusion problem mentioned in Section \ref{sec:CEFAC_ProblemFormulation} can be equivalently described as follows: setting the initial state of a normal Node $i$ to be $x_i(0)=\{X_i(0),Y_i(0)\}$ and designing a consensus algorithm that makes it so that eavesdroppers cannot infer the node's initial state and prevents attackers from interfering:
\begin{align}
	\lim\limits_{t\rightarrow\infty} x_i(t) = \sum\limits_{k = 1}^{|{\mathcal{V}}_n|} x_k(0) = \sum\limits_{k = 1}^{|{\mathcal{V}}_n|} \{X_i(0),Y_i(0)\}
	\notag
\end{align}
Since the average consensus algorithm is often used to calculate the sum of the initial states of distributed nodes, the above equation can also be equated to:
\begin{align}
	\lim\limits_{t\rightarrow\infty} x_i(t) = \frac{1}{{|{\mathcal{V}}_n|}}\sum\limits_{k = 1}^{|{\mathcal{V}}_n|} x_k(0) = \frac{1}{{|{\mathcal{V}}_n|}}\sum\limits_{k = 1}^{|{\mathcal{V}}_n|} \{X_i(0),Y_i(0)\}
	\notag
\end{align}

\subsection{Privacy-preserving WAVCCME consensus against cyberattacks}
Subsection \ref{subsec:WAVCCME_Introduce} boils down the core of the distributed evidence fusion task to the WAVCCME consensus and further transforms it to the distributed computation of KPEEV and EVE. This subsection will design two strategies, the privacy protection strategy and the attack identification-compensation strategy, to ensure that only normal nodes participate in the fusion securely.
\subsubsection{Privacy protection strategy}\label{subsec:privacyprotectstrategy}
\begin{figure}[!h]
	\centering
	\includegraphics[width=0.6\linewidth]{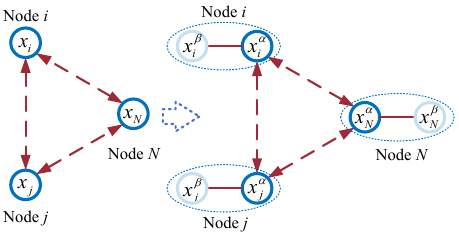}
	\caption{The privacy-preserving strategy based on state decomposition.}
	\label{fig:AttrackCEF_PrivacyStateDecomposition}
\end{figure}

Computing the KPEEV and EVE in a distributed manner requires nodes to share $X_i(0)$ and $Y_i(0)$ within their neighbors, which discloses $\boldsymbol{m}_i$. Therefore, a privacy-preserving mechanism needs to be designed. Since the privacy-preserving strategy based on state decomposition has proved to be beneficial in achieving accurate average consensus, it has been widely used in recent years \cite{calis2021aprivacy,wang2021privacy}. As shown in Fig.\ref{fig:AttrackCEF_PrivacyStateDecomposition}, this strategy decomposes the initial state $x_i(0)$ of Node $i$ into two independent random sub-states $x ^{\alpha}_i(0)$ and $x^{\beta}_i(0)$ held by Node $i$ and its virtual nodes, respectively, which satisfies $2x_i(0)=x^{\alpha}_i(0)+x^{\beta}_i(0)$ \cite{wang2019privacy}. Node $i$ exchanges information normally with its neighbors, while its virtual node exchanges information only with Node $i$. They update the state $x_i(0)$ with the following rule:
\begin{equation}
	\begin{cases}
		x_{i}^{\alpha}t + 1)=x_{i}^{\alpha}t)+\varepsilon\sum_{j\in \mathcal{N}_{i}^{-}} a_{i j}(t)\left(x_{j}^{\alpha}(t)-x_{i}^{\alpha}(t)\right)
		+\varepsilon{a}_{i,\alpha \beta}(t)\left(x_{i}^{\beta}(t)-x_{i}^{\alpha}(t)\right) \\
		x_{i}^{\beta}(t + 1)=x_{i}^{\beta}(t)+\varepsilon a_{i,\alpha \beta}(t)\left(x_{i}^{\alpha}(t)-x_{i}^{\beta}(t)\right)
	\end{cases}
	\label{eq:AttrackCEF_StateDecomposition}
\end{equation}
where $a_{i,\alpha\beta}$ is the coupling weight of $x^\alpha_i(0)$ and $x^\beta_i(0)$ and $\varepsilon\in(0,\frac{1}{\Delta}]$ is the preset parameter shared by all nodes with $\Delta=\max\{x_1,x_2 ,\cdots,x_N\}$.

However, the update rule of Eq.(\ref{eq:AttrackCEF_StateDecomposition}) is not applicable to the directed graph, and the introduction of virtual nodes poses a challenge to the identification of malicious attackers. The privacy-preserving strategy based on state decomposition and reconstruction restricts the role of virtual nodes to a finite number of iterative steps, thus balancing privacy preservation and cyberattack defense \cite{ying2023privacy}.

\begin{figure}[!h]
	\centering
	\includegraphics[width=0.6\linewidth]{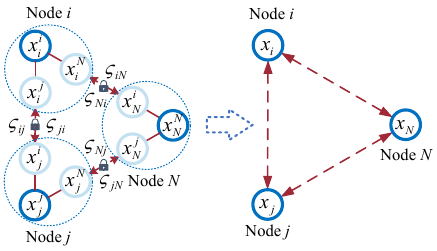}
	\caption{The state decomposition and reconstruction strategy with weight encryption.}
	\label{fig:AttackCEF_OurPrivacyScheme}
\end{figure}
To this end, a state decomposition and reconstruction strategy with weight encryption is proposed. As shown in Fig.\ref{fig:AttackCEF_OurPrivacyScheme}, this strategy decomposes the initial state $x_i(0)$ into $|{\mathcal{N}_i^+}|+1$ random sub-states $\{x_i^j(0)|j\in\{i\}\cup{\mathcal{ N}_i^+}\}$:
\begin{equation}
	x_i^i\left( 0 \right) + \sum\limits_{j \in {\mathcal{N}_i^+} } {x_i^j\left( 0 \right)}  = {x_i}\left( 0 \right)
	\label{equation:AttackCEF_OurStateDecomposition}
\end{equation}
The $x_i^j\left(0\right)$ is sent to Node $j\in{\mathcal{N}_i^+}$. In parallel, Node $i$ sends randomly generated privacy weight ${\varsigma_{ij}} \in \left[ {0,1} \right]$ encrypted by the Paillier semi-homomorphic encryption algorithm to Node $j$ for privacy protection. The Paillier is an asymmetric encryption technique that supports additive homomorphism, and its encryption and decryption are described as Algorithm \ref{alg:CEFAC_Paillier}.
\begin{algorithm}[!h]
	\caption{Paillier semi-homomorphic encryption algorithm}
	\label{alg:CEFAC_Paillier}
	\begin{algorithmic}[1]
		\REQUIRE Plaintext ${\varsigma_{ij}}$.
		\ENSURE Ciphertext $Enc({\varsigma_{ij}})$.\\
		-------------------- Key generation --------------------
		\STATE Choose sufficiently large primes $p$ and $q$ that satisfy $gcd\left(pq,\left(p\!-\!1\right)\left(q\!-\!1\right)\right)=1$;
		\STATE Get the least common multiple of $p-1$ and $q-1$: $\lambda = lcm\left(p-1,q-1\right)$;
		\STATE Generate a random number $g \in Z_{b^2}^*$, with $b=pq$;
		\STATE $\mu = \left(L\left(g^\lambda \!\!\mod b^2\right)\right)^{-1} \!\!\!\mod b$ with $L(x)=(x-1)/b$;
		\STATE The public key is $(b,g)$ and the private key is $(\lambda,\mu)$.\\
		----------------- Plaintext encryption -----------------
		\STATE A plaintext $\varsigma_{ij}$ satisfies $0\leq \varsigma_{ij} \leq b$;
		\STATE Generate a random number $r \in Z_b^*$ and $gcd(r,b) = 1$;
		\STATE Calculate the ciphertext $Enc(\varsigma_{ij}) = g^{\varsigma_{ij}}r^b \mod b^2$.\\
		---------------- Ciphertext decryption ----------------
		\STATE A ciphertext $Enc(\varsigma_{ij})$ satisfies $Enc(\varsigma_{ij}) \in Z_{b^2}^*$;
		\STATE Recover plaintext $\varsigma_{ij} = L\left(Enc(\varsigma_{ij})^\lambda \!\!\mod b^2\right)\cdot \mu \!\!\mod b$.
	\end{algorithmic}
\end{algorithm}

It should be noted that Node $i$ encrypts $\varsigma _{ij}$ using the public key of Node $j$ that is assumed to have been broadcast or configured to all nodes. The $\varsigma _{ij}$ will be private to the eavesdropper, as it does not hold the private key of Node $j$. At $t=1$, Node $i$ first decrypts the encrypted privacy weight $\{Enc\left(\varsigma _{ij}\right)|j\in {\mathcal{N}}_i^-\}$ into plaintext $\{\varsigma _{ij}|j\in {\mathcal{N}}_i^-\}$, and then updates the state according to the following rule:
\begin{equation}
	x_i(1) = x_i^i(0) + \!\!\sum\limits_{j \in {\cal N}_i^ - } \!\!{{\varsigma _{ji}}x_j^i\left( 0 \right)}  + \!\!\sum\limits_{j \in {\cal N}_i^ + } \!\!{\left( {1 - {\varsigma _{ij}}} \right)x_i^j\left( 0 \right)}
	\label{equation:AttackCEF_OurStateReconstruction}
\end{equation}
It is obvious that the sum of the node states at $t=0$ and $t=1$ is equal, i.e:
\begin{equation}
	\hspace{-0.32cm}
	\begin{aligned}
		\sum\limits_{i = 1}^N  {{x_i} \left( 1 \right)}   = & \sum\limits_{i = 1}^N    {\left(  x_i^i\left( 0 \right) +     \sum\limits_{j \in {\cal N}_i^ - }     {{\varsigma _{ji}}x_j^i \left( 0 \right)}  +     \sum\limits_{j \in {\cal N}_i^ +}    {\left( {1  -  {\varsigma _{ij}}} \right) x_i^j  \left( 0 \right)}   \right)  } \\
		=&\sum\limits_{i = 1}^N   {\left(  {x_i^i\left( 0 \right) +     \sum\limits_{j \in N_i^+ }    {x_i^j\left( 0 \right)} } \right)  }\\
		=& \sum\limits_{i = 1}^N {{x_i}\left( 0 \right)}
	\end{aligned}
\end{equation}
Therefore, the privacy-preserving strategy as in Eq.(\ref{equation:AttackCEF_OurStateDecomposition}) and Eq.(\ref{equation:AttackCEF_OurStateReconstruction}) does not impact the average consensus. The following is an in-depth analysis of the privacy-preserving performance of this strategy from two different perspectives, internal and external to the system:
\begin{enumerate}[label=(\arabic*)]
	\item The eavesdropper within the system, such as the Node $h\in\mathcal{V}_c$, obeys the rules of state decomposition and reconstruction described by Eq.(\ref{equation:AttackCEF_OurStateDecomposition}) and Eq.(\ref{equation:AttackCEF_OurStateReconstruction}). It receives partial sub-states of its incoming neighbors. The set of information that can be collected by Node $h$ at the moment $t=1$ is:
	\begin{equation}
		{\mathcal{I}}_h(1) = \{x_i^h(0),x_h^j(0),x_h^h(0),\varsigma _{hi},\varsigma _{ih}| i \in {\mathcal{N}_h^-}, j \in {\mathcal{N}_h^+}\}
		\notag
	\end{equation}
	The Node $h$ has no way of knowing $x_i(0),i\in {\mathcal{N}_h^-}$ because it fails to obtain $\{x_i^j(0),\varsigma _{ij}|j\in \{i\} \cup {\mathcal{N}_i^+}\setminus\{h\}\}$.
	\item Eavesdroppers external to the system have full access to the system topology and information transmitted between nodes, which allows them to gather more information about the initial sub-states of the nodes. The proposed privacy-preserving mechanism hides node information by encrypting weights. The initial state information about Node $i$ collected by eavesdropper $e$ at $t=1$ can be written as:
	\begin{equation}
		{\mathcal{I}}_e(1)=\{x_i^j(0),x_k^i(0),Enc(\varsigma _{ji}),Enc(\varsigma _{ij})|j \in {\mathcal{N}}_i^+, k\in {\mathcal{N}}_i^-\}
		\notag
	\end{equation}
	The eavesdropper $e$ cannot recover any of the sub-states of Node $i$ based on Eq.(\ref{equation:AttackCEF_OurStateReconstruction}) because it does not have access to the private key of Node $i$, which ensures the privacy of the initial state.
\end{enumerate}
When $t\geq{1}$, the states of all nodes have been reconstructed and fully mixed. Neither an eavesdropper inside nor outside the system can access all the initial sub-states of a particular node at the same time. Since the sum of the states of all nodes before and after the reconstruction is constant, the average consensus algorithm can be used to compute WAVCCME without worrying about the node states being recovered by any eavesdropper.

\subsubsection{Attacker identification and attack compensation strategy}
The subsection \ref{subsec:privacyprotectstrategy} has prevented the internal/external eavesdroppers in the system from inferring evidence from other nodes by reconstructing the node states. In this subsection, detection and adaptive compensation strategy is designed for DoS attacks and deception attacks to obtain the WAVCCME of normal nodes under the assumption that the directed graph $\mathcal{G}$ satisfies strong $p$-fraction robustness.

First, Node $i$ generates a state storage vector $S_i$ containing $N$ storage units, and its $j$th unit $S_{i,j}$ records the state $x_j$ of Node $j$, which are subsequently compared with each other to recognize the attack type and compute the WAVCCME. The $S_{i,j}$ is set to empty at the initial moment. The update strategy of $S_i$ is described below. For convenience, $S_i(t)$ is used here to denote the update result of $S_i$ at moment $t$.

At $t=1$, Node $i$ records the reconstructed state $x_i$ into storage unit $S_{i,i}$. From the moment $t=2$, the node starts broadcasting $S_i(t)$ to its neighbors $j\in{\mathcal{N}}_i^+$. Since there is no record to refer to or compare with, Node $i$ can only choose to trust its incoming neighbor $j\in{\mathcal{N}}_i^-$ at moment $t=2$ by updating $S_{j,j}(1)$ locally, i.e., $S_{i,j}(2)=S_{j,j}(1)$.

At $t > 2$, the node starts to recognize the type of attack and executes an appropriate defense and compensation strategy. The DoS attacker directly drops node packets. If $|\{S_j(t -1)|j \in {\mathcal{N}}_i^-\}|$ is less than $|{\mathcal{N}}_i^-|$, then it indicates that one of the incoming neighbors did not send the storage vector, i.e., a DoS attack occurs. In this case, the majority acceptance principle is adopted to avoid Node $i$ from being deceived. Specifically, if the majority storage vectors in $\{S_j(t-1)|j\in{\mathcal{N}}_i^-\}$ have the same $S_{\cdot,k}(t -1)$, Node $i$ will update $S_{i,k}(t)$ to $S_{\cdot,k}(t -1)$. And, if $|\{S_j(t -1)|j \in {\mathcal{N}}_i^-\}|$ is equal to $|{\mathcal{N}}_i^-|$, then it means that the incoming neighbor of Node $i$ has no DoS attacker. At this point, according to the $f$-fraction local cyberattacks model, Node $i$ updates $S_{\cdot,k}(t -1)$ to $S_{i,k}(t)$ is conditional on $\{S_j(t -1)|j \in {\mathcal{N}}_i^-\}$ has more than $f|{\mathcal{N}}_i^-| + 1$ storage vectors with the same $S_{\cdot,k}(t -1)$ values. Obviously, if the above update rule is observed, the node will not put the wrong $S_{\cdot,k}(t -1)$ updated to the local storage vector. Therefore, if the record $S_{l,k}(t -1)$ of the incoming neighbor $l \in {\mathcal{N}}_i^-$ is different from $S_{\cdot,k}(t -1)$, then it is a deception attacker.

To exclude the attacker's evidence from fusion, Node $i$ generates a correction storage vector $E_i$ containing $N$ cells. The $j$th cell $E_{i,j}$ of $E_{i}$ stores the set of the deception neighbors $\mathcal{V}_{D,j}\triangleq\{k|k\in{\mathcal{N}}_j^+\cup{\mathcal{N}}_j^-\cap \mathcal{V}_{D}\}$ and the set of DoS neighbors $\mathcal{V}_{DoS,j}\triangleq\{k|k\in{{\mathcal{N}}_j^+\cup{\mathcal{N}}_j^-}\cap\mathcal{V}_{DoS}\}$ identified by the Node $j\in\mathcal{V}_n$, and the corresponding correction amount $M_j$, i.e., $E_{i,j}\triangleq\{\mathcal{V}_{D,j},\mathcal{V}_{DoS,j},M_j\}$. Just like $S_i$, $E_{i,j}$ is initially set to empty, and $E_i(t)$ is used to denote the update result of $E_i$ at moment $t$. Obviously, when the state storage vectors have converged, the normal Node $j\in{\mathcal{V}_n}$ will be able to recognize the type of the neighboring attackers and thus construct $\mathcal{V}_{DoS,j}$ and $\mathcal{V}_{D,j}$.

The correction amount $M_i$ includes both information compensation and deletion. On the one hand, according to Eq.(\ref{equation:AttackCEF_OurStateReconstruction}), Node $i\in\mathcal{V}_{DoS}$ does not transmit information to Node $j\in{\mathcal{N}}_i^+$, while Node $l\in{\mathcal{N}}_i^-$ sent to Node $i$ will also not be updated to $x_i(1)$. Therefore, the missing information ${\varsigma_{il}}x_l^i(0)$ for Node $l\in{\mathcal{N}}_i^-$ must be compensated. On the other hand, the correction against the attacker $i\in\mathcal{V}_{D}$ should not only compensate for the missing information decomposed from the Node $j\in{\mathcal{N}}_i^-$, but also weed out the erroneous information added to the Node $j\in{\mathcal{N}}_i^+$. Therefore, there is:
\begin{equation}
	M_j=L_j-W_j=\sum\limits_{k\in{\mathcal{N}}_j^+\cap\mathcal{V}_c}{\varsigma_{jk}}x_j^k(0) - \sum\limits_{k\in{\mathcal{N}}_j^-\cap\mathcal{V}_{D}}{\varsigma_{kj}}x_k^j(0)
\end{equation}
A normal node neither receives correction storage vectors from neighboring attackers nor sends correction storage vectors to any neighboring attackers. In the update strategy, $E_i$ is the same as $S_i$ to ensure that the cyberattack does not contaminate the correction amount. Therefore, when the correction storage vector is consensualized to the whole network, all normal nodes get the identities of the attackers and the corresponding corrections.

In the consensus above, normal nodes update their status using the following rules:
\begin{equation}
	{x_i}(t) = {\varepsilon _i}{x_i}(t - 1) + (1 - {\varepsilon _i})\left( {\frac{{\sum\limits_{j \in {{\mathcal{V}}_{n,i}}\left( t \right)} S_{i,j}\left(t\right) + \sum\limits_{j \notin {{\mathcal{V}}_{n,i}}\left( t \right)}M_j\left( t \right) }}{{\left| {{{\mathcal{V}}_{n,i}}\left( t \right)} \right|}}} \right)
	\label{eq:AttackCEF_Update}
\end{equation}
where $\varepsilon_i \in [0,1)$ is the control gain, and ${{\mathcal{V}}_{n,i}}\left(t\right)$ denotes the set of normal nodes known to Node $i$ at moment $t$:
\begin{equation}
	{{\mathcal{V}}_{n,i}}\left( t \right) = {\mathcal{V}}\subseteq \left( {\mathop  \cup \limits_{{E_{i,j}}\left( t \right) \ne \emptyset } \left( {{{\mathcal{V}}_{DoS,j}}\left( t \right) \cup {{\mathcal{V}}_{D,j}}\left( t \right)} \right)} \right)
	\label{eq:AttackCEF_UpdateNormalNodes}
\end{equation}
\begin{thm}
	\label{theorem_3}
	Under the update rule Eq.(\ref{eq:AttackCEF_Update}), the states of all normal nodes will converge to:
	\begin{equation}
		\lim\limits_{t\rightarrow\infty}x_i(t) = \lim\limits_{t\rightarrow\infty}\{X_i(t),Y_i(t)\} =  \frac{1}{|{\mathcal{V}}_{n}|}\sum_{k\in{\mathcal{V}}_{n}} \{X_k(0),Y_k(0)\}
		\label{eq:AttackCEF_ConvergencyValue}
	\end{equation}
\end{thm}
\begin{pf}
	According to Theorem 2 and Theorem 3 in \cite{ying2023privacy}, if the graph $\mathcal{G}$ satisfies the $f$-fraction local cyberattacks model and $p$-fraction robustness with $2f<p\leqslant{1}$, then the storage unit of normal nodes, $S_{i,j}(t)$, will eventually converge to $x_j(1)$. As the attack is persistent and of a constant type, the Node $i\in{\mathcal{V}}_n$ is evidently able to accurately recognize the attacker's identity and type by employing the above detection strategy to construct ${\mathcal{V}}_{D,i}$ and ${\mathcal{V}}_{DoS,i}$ and write them to $E_{i,i}$. Thus, as $E_i(t)$ is updated, Eq.(\ref{eq:AttackCEF_UpdateNormalNodes}) will converge to:
	\begin{equation}
		\lim\limits_{t\rightarrow\infty}{{\mathcal{V}}_{n,i}}\left( t \right) ={\mathcal{V}}_{n}
	\end{equation}
	And then there is:
	\begin{equation}
		\begin{aligned}
			&\lim\limits_{t\rightarrow\infty}\!\!{\sum\limits_{j\in{{\mathcal{V}}_{n,i}}\left(t\right)}\!\!\!S_{i,j}\left(t\right)+\!\!\!\!\!\sum\limits_{j\notin{{\mathcal{V}}_{n,i}}\left(t\right)}\!\!\!M_j\left(t\right)}\\
			=&\sum\limits_{i\in{\mathcal{V}}_n} \left({x_i(1)} + \!\!\!\!{\sum\limits_{j\in{\mathcal{V}}_c\cap{\mathcal{N}}_i^+}\!\!\!\!\!{\varsigma_{ij}}x_i^j(0)}-\!\!\!\!{\sum\limits_{j\in{\mathcal{V}}_D\cap{\mathcal{N}}_i^-}\!\!\!\!\!{\varsigma_{ji}}x_j^i(0)}\right) \\ 
			=&\sum\limits_{i\in{\mathcal{V}}_n} \left(x_i^i(0) + 
			\sum\limits_{j\in{\mathcal{N}}_i^-}\!{{\varsigma_{ji}}x_j^i(0)}-\!\!\!\!\!
			\sum\limits_{j\in{\mathcal{V}}_D\cap {\mathcal{N}}_i^-} \!\!\!\!\!{\varsigma_{ji}}x_j^i(0)+\!\!\!\!\!
			\sum\limits_{j\in{\mathcal{V}}_c\cap{\mathcal{N}}_i^+} \!\!\!\!\! {\varsigma _{ij}}x_i^j(0) +\!\!\!\sum\limits_{j\in{\mathcal{N}}_i^+}(1-\varsigma_{ij})x_i^j(0) \right)  \\
			=&\sum\limits_{i\in{{\mathcal{V}}_n}} \left( x_i^i(0) + \sum\limits_{j\in{\mathcal{N}}_i^+}\,\,\,\,\,x_i^j(0)+\!\!\!\!\!\!\!
			\sum\limits_{\scriptstyle j\in{\mathcal{N}}_i^- \atop\scriptstyle j\in{{\mathcal{V}}_n}\cup{{\mathcal{V}}_{DoS}}} \!\!\!\!\!\!\!{{\varsigma_{ji}}x_j^i(0)} - \!\!\!\sum\limits_{j\in{\mathcal{N}}_i^+} {\varsigma _{ij}}x_i^j(0) +\!\!\!\!\!\sum\limits_{j\in{\mathcal{V}}_c\cap{\mathcal{N}}_i^+}\!\!\!\!{\varsigma_{ij}}x_i^j(0) \right)  \\ 
			=& \sum\limits_{i \in {{\mathcal{V}}_n}} \left( x_i(0) + \!\!\!\! \sum\limits_{j\in{\mathcal{V}}_n  \cap {\mathcal{N}}_i^-} \!\!\!\!\!{{\varsigma _{ji}}x_j^i(0)}  - \!\!\!\!\sum\limits_{j \in {{\mathcal{V}}_n} \cap {\mathcal{N}}_i^+}  \!\!\!\!{\varsigma _{ij}}x_i^j\left( 0 \right) \right)\\ 
			=& \sum\limits_{i \in {{\mathcal{V}}_n}} {{x_i}(0)}
		\end{aligned}
	\end{equation}
	Since the Eq.(\ref{eq:AttackCEF_Update}) can be written as:
	\begin{equation}
		{{x_i}(t) - \frac{{\sum\limits_{j \in {{\mathcal{V}}_{n,i}}\left( t \right)}\!\!\!\!\! S_{i,j}\left(t\right) \!+\!\!\!\!\! \sum\limits_{j \notin {{\mathcal{V}}_{n,i}}\left( t \right)}\!\!\!\!\!M_j\left( t \right) }}{{\left| {{{\mathcal{V}}_{n,i}}\left( t \right)} \right|}}}  = {\varepsilon _i}\left( {{x_i}(t - 1) - \frac{{\sum\limits_{j \in {{\mathcal{V}}_{n,i}}\left( t \right)}\!\!\!\!\! S_{i,j}\left(t\right) \!+\!\!\!\!\! \sum\limits_{j \notin {{\mathcal{V}}_{n,i}}\left( t \right)}\!\!\!\!\!M_j\left( t \right) }}{{\left| {{{\mathcal{V}}_{n,i}}\left( t \right)} \right|}}} \right)
	\end{equation}
	Thus, when $0\leqslant{\varepsilon _i}<1$, the system is Schur-stable and:
	\begin{equation}
		\lim\limits_{t\rightarrow\infty} x_i(t) = \frac{\sum_{k \in {{\mathcal{V}}_n}} {{x_k}(0)}}{|{\mathcal{V}}_n|}=\frac{1}{|{\mathcal{V}}_{n}|}\!\sum_{k\in{\mathcal{V}}_{n}} \left\{X_k(0),Y_k(0)\right\} 
	\end{equation}
\end{pf}
At this point, all normal nodes could obtain $\bar{X}\!\!=\!\!\sum_{i\in\mathcal{V}_n}\!X_i(0)$ and $\bar{Y}\!\!=\!\!\sum_{i\in\mathcal{V}_n}\!Y_i(0)$, and further get $\boldsymbol{m}_{avg|A}\!=\!\textbf{1}^T\text{diag}(\bar{Y})^{-1}\bar{X}$.
\begin{algorithm}[!h]
	\caption{The WAVCCME consensus with evidence privacy-preserving under cyberattacks}
	\label{alg:AttackCEF_ConditionalAverageConsensus}
	\begin{algorithmic}[1]
		\REQUIRE Evidence $\boldsymbol{m}_1,\boldsymbol{m}_2, \cdots ,\boldsymbol{m}_N$, control gain $\varepsilon_1,\varepsilon_2,\cdots,\varepsilon_N$, FoD $\Omega = \{ {{\hat A}_1},{{\hat A}_2}, \cdots ,{{\hat A}_n}\} $, number of step for iteration termination $N_s$.
		\ENSURE The WAVCCME of $|\mathcal{V}_n|$ normal nodes $\boldsymbol{m}_{avg|A}$.
		\STATE ${t}\leftarrow{0}$;
		\FOR {Node ${i}\in[1,N]$, in parallel}
		\WHILE{True}
		\IF{$t=0$}\label{step:Alg_wawvvccme_prepare_Start}
		\STATE Calculate the difference measure between $\boldsymbol{m}_i$ and $n$ pieces of event evidence;
		\STATE Calculate the support of $\boldsymbol{m}_i$ for $n$ events $\{sup_{ki}|k=1,2,\cdots,n\}$;
		\STATE Construct and decompose $x_i(0)$ into $\{x_i^j(0)|j\in{\mathcal{N}}_i^+\cap\{i\}$;
		\STATE Generate privacy weights $\{\varsigma _{ij}|j\in{\mathcal{N}}_i^+\cap\{i\}\}$;
		\ENDIF\label{step:Alg_wawvvccme_prepare_End}
		\IF{$t =1$}
		\STATE Encrypt $\varsigma_{ij}$ with the public key of Node $j\in{\mathcal{N}}_i^+$ and send it to Node $j$ along with $x_i^j(0)$;
		\STATE Decrypt the ciphertext of the Node $j\in{\mathcal{N}}_i^-$ with its own private key to obtain $\varsigma _{ij}$;
		\STATE Reconstruct the state by Eq.(\ref{equation:AttackCEF_OurStateReconstruction}) to get $x_i(1)$;
		\STATE Generate $S_{i}(1)$ and store $x_i(1)$ into $S_{i,i}(1)$;
		\ENDIF
		\IF{$t=2$}
		\STATE Send $S_{i}(1)$ to Node $j\in{\mathcal{N}}_i^+$ and receive $\{S_{j}(1)|j\in{\mathcal{N}}_i^-\}$;
		\STATE $\forall j\in {\mathcal{N}}_i^-,S_{i,j}(2) = S_{j,j}(1)$;
		\ENDIF
		\IF{$t>2$}
		\STATE Send $S_{i}(t-1)$ to Node $j\in {\mathcal{N}}_i^+$;
		\IF{The number of received $S_j(t-1)$ is less than $|{\mathcal{N}}_i^-|$}
		\STATE $S_{i,k}(t)=S_{\cdot,k}(t-1)$ if the majority in $\{S_{j}(t\! -\!1)|j\in {\mathcal{N}}_i^-\}$ have the same record $S_{\cdot,k}(t-1)$;
		\ELSE
		\STATE When there are more than $f|{\mathcal{N}}_i^-|+1$ storage vectors in $\{S_{j}(t-1)|j\in{\mathcal{N}}_i^-\}$ have the same $S_{\cdot,k}(t-1)$, then $S_{i,k}(t)=S_{\cdot,k}(t-1)$;
		\ENDIF
		\IF {$\{S_{j}(t-1)|j\in {\mathcal{N}}_i^-\}$ does not change in more than $N_s$ steps}
		\STATE Node $i$ constructs ${\mathcal{V}}_{DoS,i}$,${\mathcal{V}}_{D,i}$,$M_i$ and stores them to $E_{i,i}(t)$.
		\STATE Broadcast and update $E_{i}(t)$ among normal nodes.
		\STATE Terminate the loop when $x_i(t)$ does not change for more than $N_s$ steps;
		\ENDIF
		\STATE Update $x_i(t)$ by Eq.(\ref{eq:AttackCEF_Update});
		\ENDIF
		\STATE ${t}\leftarrow{t+1}$;
		\ENDWHILE
		\STATE $\boldsymbol{m}_{avg|A}\leftarrow \textbf{1}^T\text{diag}(\bar{Y})^{-1}\bar{X}$;
		\ENDFOR
	\end{algorithmic}
\end{algorithm}

The Algorithm \ref{alg:AttackCEF_ConditionalAverageConsensus} shows the pseudo-code for computing the WAVCCME of normal nodes in a distributed system suffering from cyberattacks and protecting the privacy of evidence. At $t=0$, Node $i$ constructs pieces of event evidence corresponding to $n$ mutually exclusive events, and then calculates the support of $\boldsymbol{m}_i$ for these events $\{sup_{ki}|k=1,2,\cdots,n\}$. Subsequently, the initial state $x_i(0)$ of Node $i$ is constructed according to Eq.(\ref{eq:AttrackCEF_YiDef})-(\ref{eq:AttrackCEF_XiDef}).

At $t=1$, the nodes mix neighboring sub-states into their own state by decomposition and reconstruction for privacy protection. Given that the Paillier homomorphic encryption algorithm is primarily designed for integer encryption, the $\varsigma_{ij}$ needs to be scaled appropriately before it is encrypted. Specifically, this work intercepts the integer portion of $\varsigma_{ij}$ multiplied by $10^4$ to be used for encryption to ensure that it retains sufficient precision before and after encryption. For the case of Node $i$ encrypting $\varsigma _{ij}$ using the public key of Node $j$, this work defaults to the fact that the public keys of all nodes have been broadcast to the entire network in advance. After receiving the privacy weight, Node $j$ decrypts the data using its own private key and divides the decrypted privacy weight by $10^4$ to obtain $\varsigma _{ij}$, which guarantees the security of the data transmission and the privacy of the node's evidence.

Subsequently, the pseudo-code shows how the mixed state $x_i(1)$ is stored into the vector $S_i$ and completes the propagation across the network. In this process, the resilient consensus mechanism is used to defend against possible cyberattacks and to recognize the identity and type of the attacker. After the consensus of $S_i$ is reached, normal nodes can mark the identity of the attacker and generate the corresponding correction amount, which in turn realizes the WAVCCME consensus.

\subsection{Obtaining credible fusion result locally at the node}
In this subsection, we illustrate in detail how nodes obtain the credible fusion result of normal nodes evidence based on $\boldsymbol{m}_{avg|A}$ independently.
\begin{algorithm}[!h]
	\caption{CEFAC}
	\label{alg:AttackCEF_CEF}
	\begin{algorithmic}[1]
		\REQUIRE Evidence $\boldsymbol{m}_1,\boldsymbol{m}_2, \cdots ,\boldsymbol{m}_N$, iteration termination threshold $\delta$, distance coefficient $\tau$, FoD $\Omega = \{ {{\hat A}_1},{{\hat A}_2}, \cdots ,{{\hat A}_n}\}$.
		\ENSURE Fusion result $\boldsymbol{m}_{\oplus}$.
		\STATE The nodes obtain the $\boldsymbol{m}_{avg|A}$ of normal nodes in a distributed manner utilizing the Algorithm \ref{alg:AttackCEF_ConditionalAverageConsensus};
		\STATE $t \leftarrow 1$;
		\FOR {Node $i \in [1,N]$, in parallel}\label{step:alg_ACEF_Fusion_Start}
		\STATE Initialize the event prior probability $P^{(0)}(A)$ by Eq.(\ref{eq:AttackCEF_AvgEventProb});
		\WHILE{True}
		\STATE Calculate $\boldsymbol{m}^{(t)}_{avg}$ by Eq.(\ref{eq:AttackCEF_GetAvgMass});
		\STATE Obtain the fusion result $\boldsymbol{m}^{(t)}_{\oplus}$ according to Eq.(\ref{eq:Introduction_DempsterCombinationRule}) and Eq.(\ref{eq:ICEF_Summary_3});
		\STATE Calculate the event probability $P^{(t)}(A)$ according to Eq.(\ref{eq:ICEF_Summary_3}) and Eq.(\ref{eq:Introduction_MassToBetP}); 
		\IF{$\|P^{(t)}(A)-P^{(t-1)}(A)\|_2 \leq \delta $}
		\STATE Jump out of the loop;
		\ENDIF
		\STATE $t \leftarrow t+1$;
		\ENDWHILE
		\STATE $\boldsymbol{m}_{\oplus} \leftarrow \boldsymbol{m}_{\oplus}^{(t)}$.
		\ENDFOR\label{step:alg_ACEF_Fusion_End}
	\end{algorithmic}
\end{algorithm}

As shown in Algorithm \ref{alg:AttackCEF_CEF}, the nodes first collaboratively compute the WAVCCME $\boldsymbol{m}_{avg|A}$ of all normal nodes according to Algorithm \ref{alg:AttackCEF_ConditionalAverageConsensus}. Then, each node independently obtains the credible fusion $\boldsymbol{m}_{avg|A}$ based on $\boldsymbol{m}_{avg|A}$ for $|\mathcal{V}_n|$ pieces of evidence. Here the fusion is done in an iterative manner because the process involves complex operations such as the Dempster operator. And the initial prior probabilities of events in the FoD are assigned on average to start the iteration, since it is ignorance about these events at the beginning:
\begin{equation}
	\begin{aligned}
		P^{(0)}(A)&\triangleq \left[p(\hat A_1),p(\hat A_2),\cdots,p(\hat A_n)\right]^T \\
		&= \left[\frac{1}{N},\frac{1}{N},\cdots,\frac{1}{N}\right]^T
	\end{aligned}
	\label{eq:AttackCEF_AvgEventProb}
\end{equation}
where $p(\hat A_i)$ is the a priori probability of the event $\hat A_i \in \Omega$, and $P^{(0)}(A)$ is a column vector consisting of the a priori probabilities of $n$ events at $t=0$. At iteration step $t$, the node computes $\boldsymbol{m}_{avg|A}$ based on the event prior probabilities $P^{(t)}(A)$ and $\boldsymbol{m}_{avg|A}$:
\begin{equation}
	\boldsymbol{m}^{(t)}_{avg} = \boldsymbol{m}_{avg|A} P^{(t-1)}(A)
	\label{eq:AttackCEF_GetAvgMass}
\end{equation}
The fusion result $\boldsymbol{m}_{\oplus}^{(t)}$ is obtained by fusing $\boldsymbol{m}_{avg}^{(t)}$ for $|\mathcal{N}_n|-1$ times. Then, the $BetP$ function is used to convert $\boldsymbol{m}_{\oplus}^{(t)}$ into a prior probability $P^{(t)}(A)$ for the next iteration. This iteration is executed until $\boldsymbol{m}_{\oplus}^{(t)}$, $P^{(t)}(A)$, and $\boldsymbol{m}_{avg|A}$ converge.

\section{Simulation and analysis}
\label{sec:CEFAC_Simulation}
\subsection{Test on the fusion of high-conflict evidence}
Addressing the counterintuitive fusion result triggered by fusing high-conflict evidence is the primary problem for multi-source evidence fusion. To this end, this subsection takes the simulation scenario given by \cite{denoeux2021distributed} as an example to test the effectiveness of CEFAC in solving the high conflict problem. It should be noted that this test does not consider cyberattacks, so instead of using the Algorithm \ref{alg:AttackCEF_ConditionalAverageConsensus} to defend against eavesdroppers and attackers, the linear average consensus algorithm \cite{denoeux2021distributed} is used to compute all the nodes' WAVCCME, and the fusion result is obtained according to the lines \ref{step:alg_ACEF_Fusion_Start}-\ref{step:alg_ACEF_Fusion_End} of Algorithm \ref{alg:AttackCEF_CEF}.

\begin{figure}[!h]
	\centering
	\includegraphics[width=0.65\linewidth]{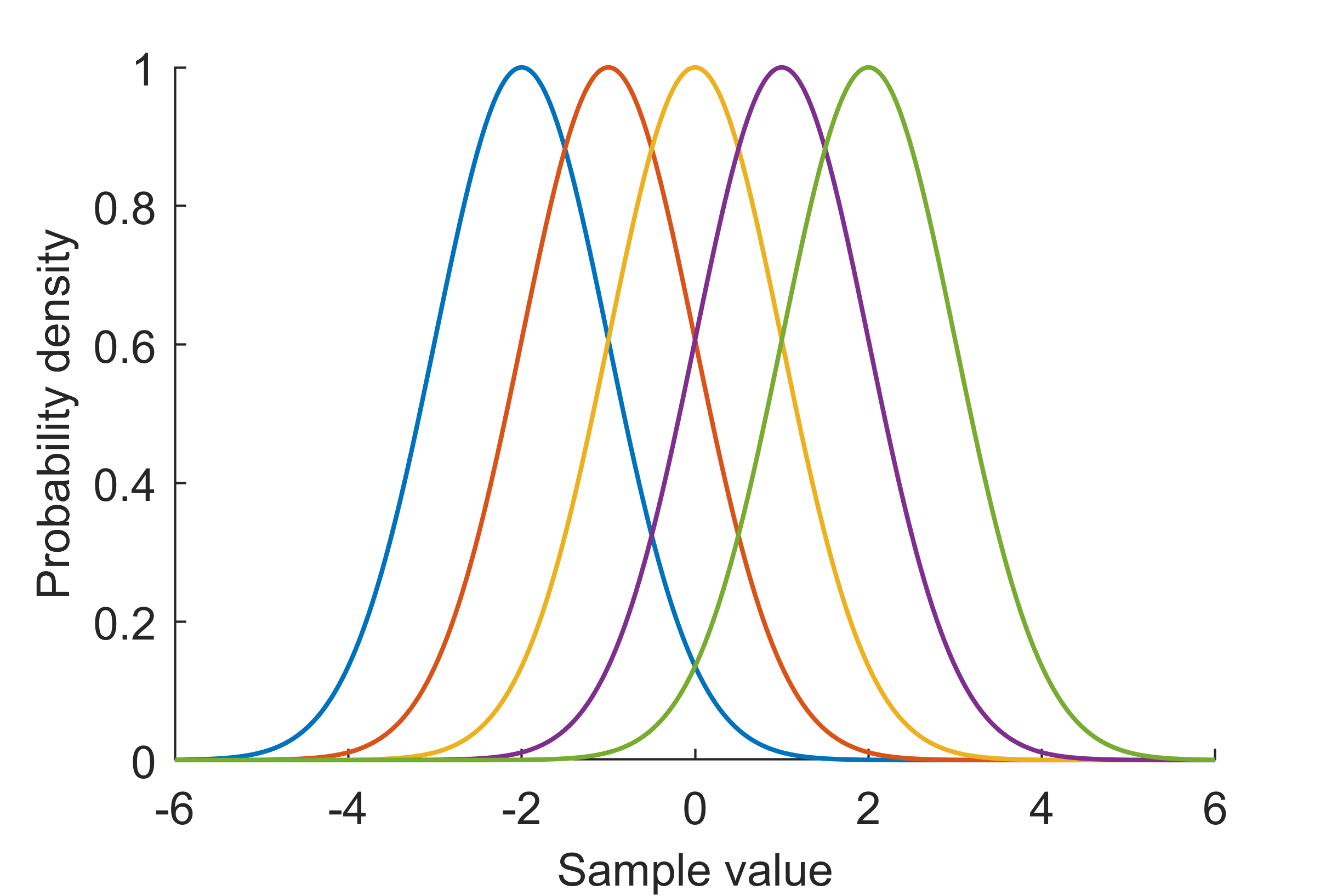}
	\caption{Probability density curves for the five categories}
	\label{fig:AttackCEF_FiveCategoryCurve}
\end{figure}
The test scenario is a classification task involving five categories $\Omega = \{ \hat{A}_1, \hat{A}_2, \hat{A}_3, \hat{A}_4,\hat{A}_5\}$. As shown in Fig.\ref{fig:AttackCEF_FiveCategoryCurve}, the probability density curves of the five categories are all normal distributions with variance 1. Corresponding to the categories $\{\hat{A}_1\}$, $\{\hat{A}_2\}$, $\{\hat{A}_3\}$, $\{\hat{A}_4\}$, and $\{\hat{A}_5\}$, the probability density functions have mean values of -2, -1, 0, 1, and 2, respectively.

The node communication network is set up as a strongly connected undirected graph consisting of 20 nodes with a connection density of 0.4. These nodes independently observe the target $T$ of category $\{\hat{A}_1\}$, with 15 nodes reporting observations obeying the normal distribution $N(-2,1)$, and the other 5 nodes sampling observations from the normal distribution $N(2,1)$. To recognize the category of $T$, the 20 nodes first use BKNN \cite{liu2013anew} as a base classifier to convert their own observations into evidence $\boldsymbol{m}_1\sim\boldsymbol{m}_{20}$ defined on $\Omega$. The classifier parameters are set to $\gamma_{t_a} = 2$, $\gamma_{t_r} = 5$, $K = 40$, and $N_s = 100$.

Since the probability density functions of the five categories have a high cross-entropy, categorizing targets based on observations can easily lead to indistinguishability or misclassification when the number of samples is small. In other words, when observations are transformed into evidence, there may be evidence fusion results more in favor of the third category or no high-conflicting evidence. To satisfy the test requirements, i.e., the category of the target $T$ is $\{\hat{A}_1\}$ and there is high conflict among the evidence, the sampling results are filtered here using the traditional centralized credible fusion (CCEF) method, for example, the method mentioned in \cite{liu2011combination}. Specifically, 15 and 5 sample points are first randomly sampled from category $\{\hat{A}_1\}$ and category $\{\hat{A}_5\}$ as node observations, respectively. Then, the BKNN classifier is used to convert the 20 observations into 20 pieces of evidence. After that, the 20 pieces of evidence are fused using CCEF and Dempster combination rules, respectively. When the CCEF recognizes the target belongs to the category $\{\hat{A}_1\}$ and the fusion result of the Dempster combination rule recognizes the target does not belong to the category $\{\hat{A}_1\}$, the group of samples is accepted; otherwise, the group of samples is rejected.

This test compares CEFAC and two distributed evidence fusion algorithms, COF-Based and RANSAC-Based, and adopts the method proposed in the literature \cite{liu2011combination} as the CCEF benchmarking method for sample selection. The parameters of the three distributed fusion algorithms are set as follows:
\begin{enumerate}[label=(\arabic*)]
	\item CEFAC: Calculate the WAVCCME of all nodes using linear average consensus and follow Algorithm \ref{alg:AttackCEF_CEF} lines \ref{step:alg_ACEF_Fusion_Start}-\ref{step:alg_ACEF_Fusion_End} obtain evidence fusion result.
	\item RANSAC-Based: According to \cite{denoeux2021distributed}, the parameters of this method are set as follows: the size of the random subsample $\nu=5$, the probability of success $p_{suc}=0.9999$, the probability of inlier point $p_{in}=0.9$, and the conflict threshold $\tau=0.5$ (the suggested range of values for $\tau$ is $[0.4,0.6]$).
	\item COF-Based: According to \cite{zhao2023information}, the parameters of this method are set as follows: the COF threshold $\tau_{COF}=0.5$ (suggested value range is $[0.4,0.9]$) and distance threshold $\tau_{dist}=0.7$ (the suggested value range is $[0.4,0.7]$).
\end{enumerate}

\begin{figure*}[!h]
	\centering
	\includegraphics[width=1\linewidth]{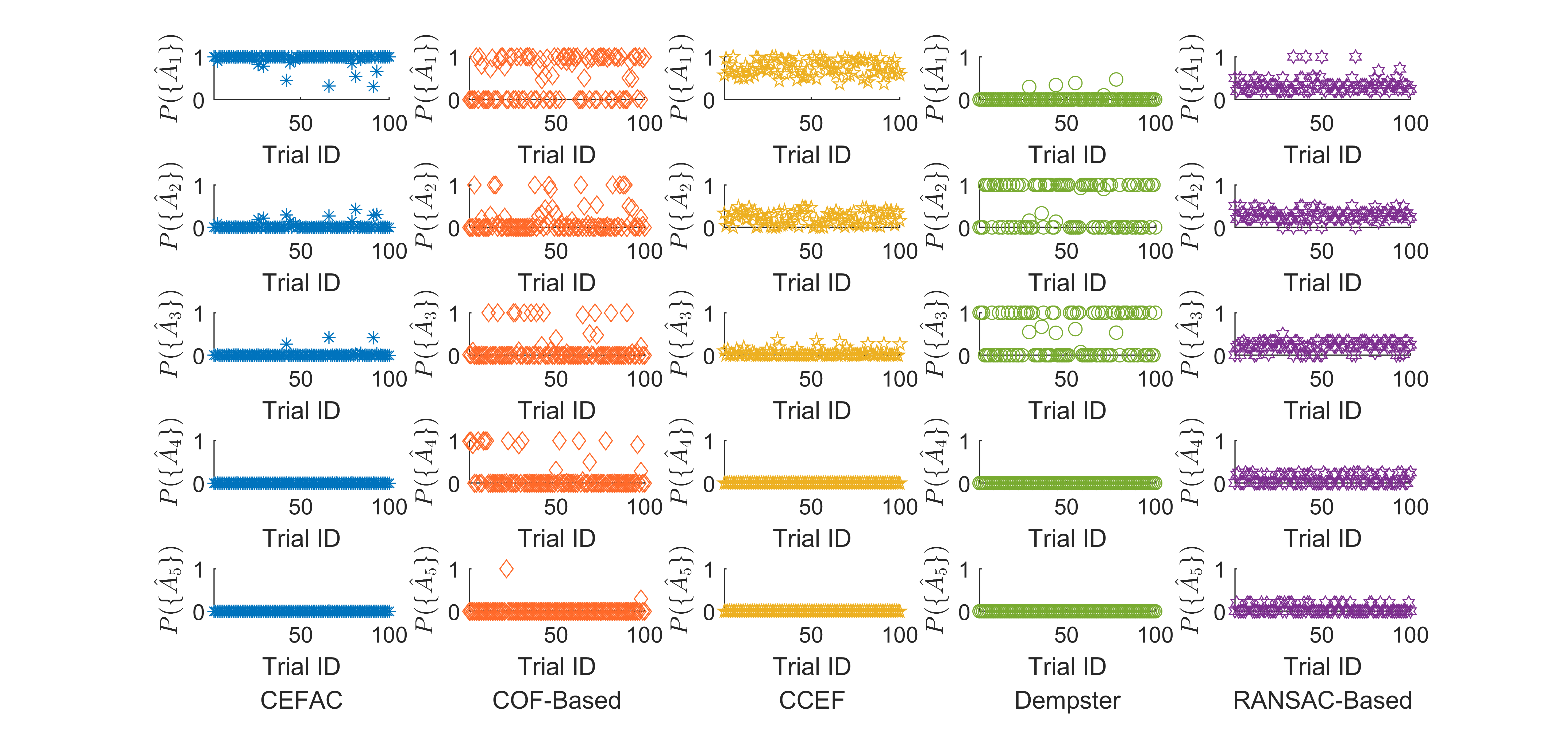}
	\caption{The pignistic probabilities of the fusion results for the five fusion methods.}
	\label{fig:AttackCEF_FusionResPignistic}
\end{figure*}

\begin{figure}[!h]
	\centering
	\includegraphics[width=0.7\linewidth]{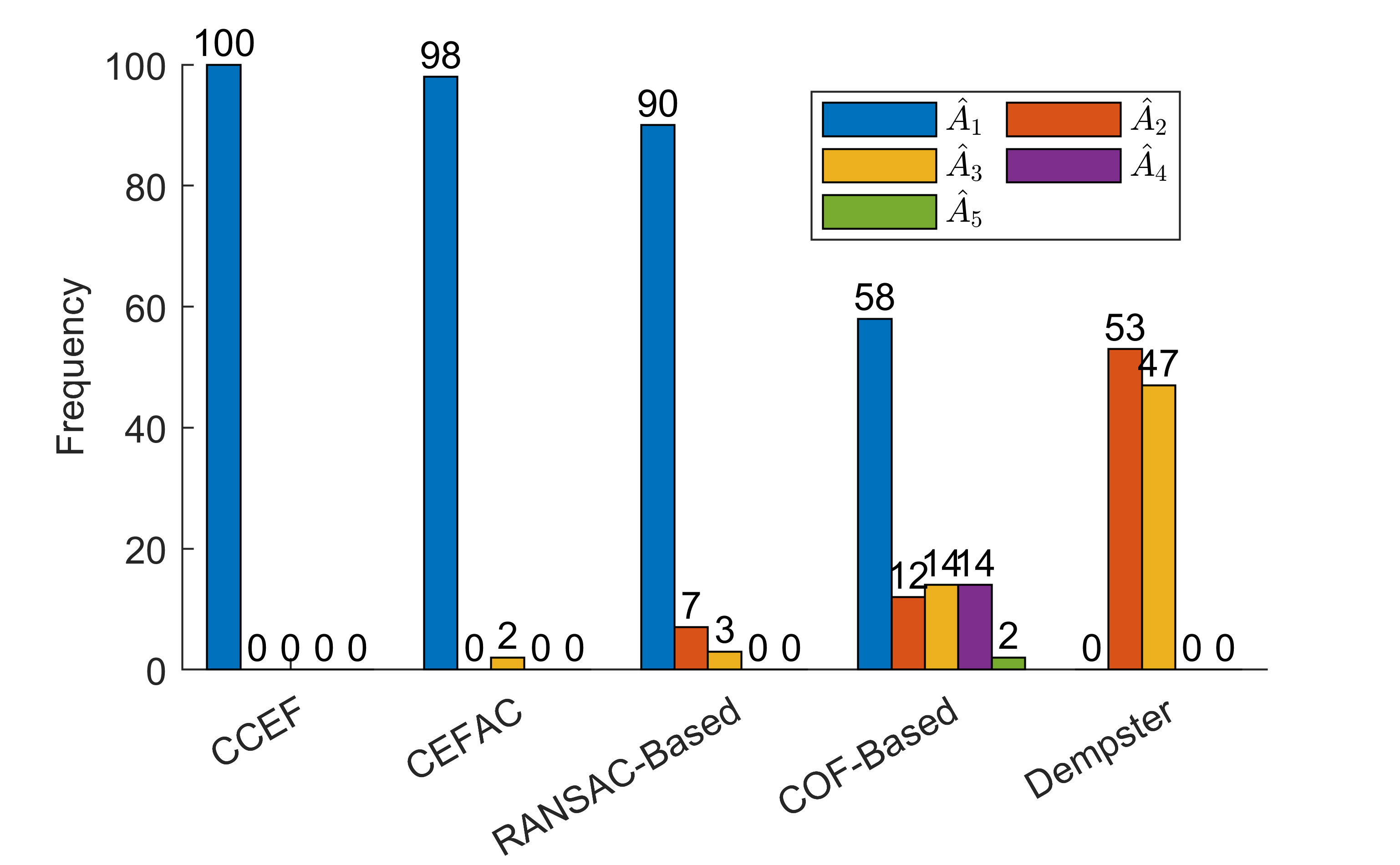}
	\caption{Statistics on the frequency of recognition results of different fusion methods.}
	\label{fig:AttackCEF_ClassficationAccuracy}
\end{figure}
Fig.\ref{fig:AttackCEF_FusionResPignistic} and Fig.\ref{fig:AttackCEF_ClassficationAccuracy} show the Pignistic probability and the frequency of classification results for the five evidence fusion methods in 100 Monte Carlo trials, respectively. It is observed that the CCEF accurately recognizes the target $T$ as the category $\hat{A}_1$ in all trials, while the Dempster combination rule recognizes the target $T$ as the category $\hat{A}_2$ or $\hat{A}_3$. It means that all the experiments fulfill the requirement of testing the fusion algorithm's ability to resist high conflicts.

The CEFAC obtains the highest recognition accuracy among the three distributed evidence fusion algorithms. Out of 100 trials, the CEFAC has 98 correct recognitions, which is higher than the 90 of RANSAC-Based and the 58 of COF-Based. In terms of Pignistic probabilities, COF-Based generally has less confidence than CEFAC and RANSAC-Based for category $\hat{A}_1$. The CEFAC performs the best of the three methods, supporting the category $\hat{A}_1$ with close to 1 confidence in the majority of cases. The RANSAC-Based, on the other hand, supports the categories $\hat{A}_2$, $\hat{A}_3$, and $\hat{A}_4$ with a probability close to 1 in a significant number of trials. As mentioned in the introduction, RANSAC-Based and COF-Based methods use random sampling or fusion of neighborhood evidence as a benchmark for normal/disturbing evidence discrimination. When highly conflicting evidence exists, the benchmark will inevitably appear counterintuitive, thus triggering wrong decisions. In addition, these two methods obtain the final fusion results by discarding a significant portion of the sources' evidence as completely wrong, which leads to a loss of information. This is also an important factor that leads to a decrease in the confidence level of the fusion results.

\begin{figure*}[!h]
	\centering
	\includegraphics[width=1\linewidth]{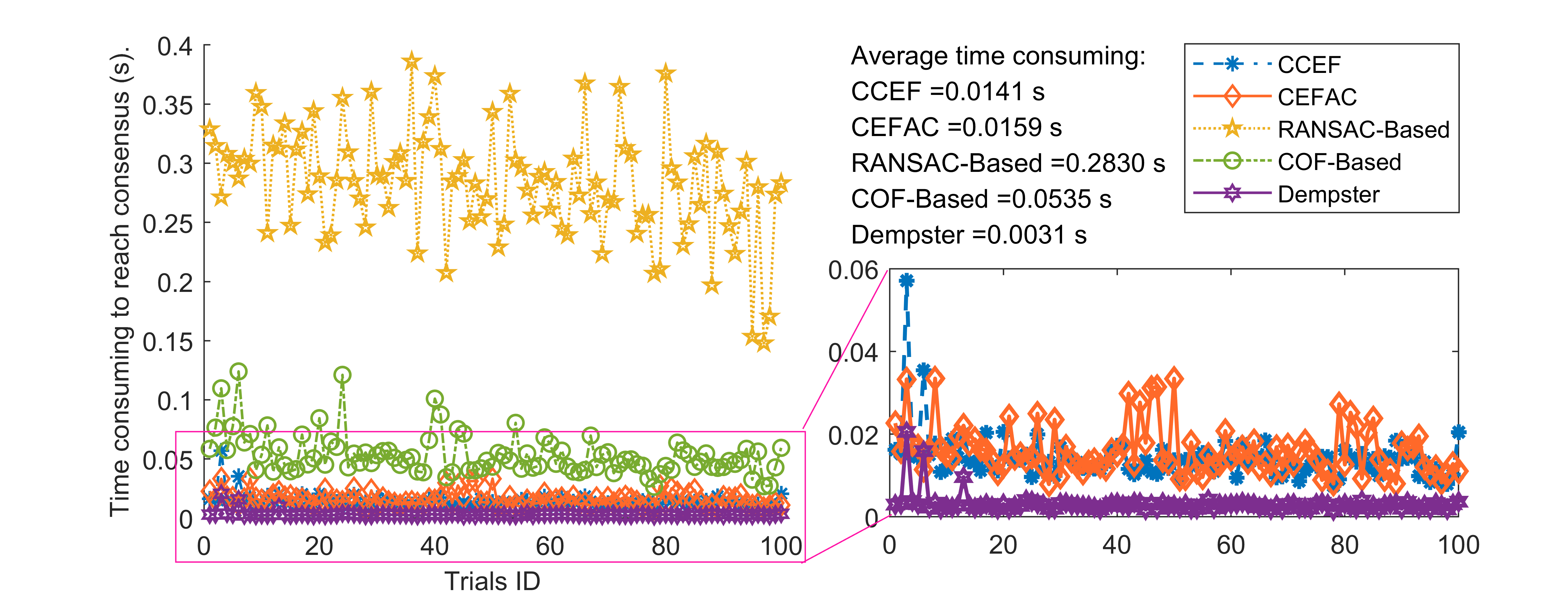}
	\caption{Time-consuming for different fusion methods}
	\label{fig:AttackCEF_TimesConsuming}
\end{figure*}
By compressing distributed evidence credibility fusion into a singular and simple consensus, the CEFAC not only implicitly solves the challenge of evaluating the credibility of multi-source evidence in a distributed system but also exhibits faster runtimes than the RANSAC-Based and the COF-Based. The runtimes of the five methods over 100 trials are given in Fig.\ref{fig:AttackCEF_TimesConsuming}. For the centralized fusion methods, including CCEF and Dempster, the runtime is defined as the total time from algorithm start to completion. For the three distributed fusion methods, the runtime, on the other hand, is defined as the average runtime per node, i.e., the total duration of the algorithm from initiation to termination divided by the total number of nodes in the network. It is shown that CEFAC has the lowest runtime among all distributed methods, even close to CCEF.

In summary, the credible fusion of distributed evidence by consensus WAVCCME offers both accuracy and speed compared to existing algorithms, so it is expected to be widely used in real-world scenarios and provides a solid foundation for further research and practice of distributed evidence fusion. In addition, since the core of fusion is simplified as a distributed summation problem, the proposed algorithm can greatly reduce the resource consumption and is well adapted to networks with an arbitrary number of nodes.

\subsection{Test on resistance to cyberattacks}

\begin{figure}[!h]
	\centering
	\includegraphics[width=0.6\linewidth]{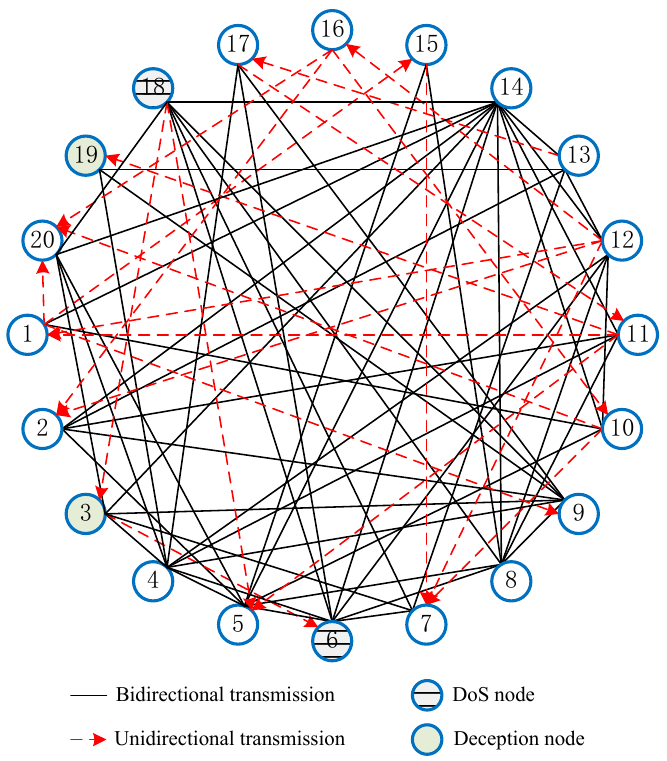}
	\caption{Directed communication network with 20 nodes}
	\label{fig:AttackCEF_DigraphOf20Nodes}
\end{figure}
This test uses a distributed collaborative identification task to evaluate the performance of CEFAC under cyberattacks. Fig.\ref{fig:AttackCEF_DigraphOf20Nodes} demonstrates a distributed reconnaissance system consisting of 20 intelligent unmanned platforms. These platforms are deployed in forward positions to build a defense system that enables the identification of enemy targets in order to provide early warning of potential attacks. Specifically, the reconnaissance system categorizes battlefield targets into three categories: unmanned aerial vehicles (UAVs), land combat vehicles (LCVs), and personnel, denoted as $\Omega = \{\hat{A}_1,\hat{A}_2,\hat{A}_3\}$. Depending on their respective deployment areas and platform characteristics, the platforms are fitted with reconnaissance equipment with different functions. The equipment complements each other and operates in synergy to provide essential support for the realization of flexible, efficient, and cost-effective battlefield reconnaissance and identification missions.

Considering multiple factors such as the command and control relationship, the performance characteristics of each unmanned platform, and the battlefield survival requirements, the communication relationships among the 20 unmanned combat platforms are abstracted into a directed transmission network as shown in Fig.\ref{fig:AttackCEF_DigraphOf20Nodes}. The black solid line segment in the figure indicates the two-way communication relationship between platforms, while the red dashed arrow indicates the one-way reporting relationship between platforms. In order to avoid its operational intent being discovered in advance, the enemy implements a variety of soft-kill means on our unmanned combat platforms, such as electromagnetic jamming, infrared deception, and cyberattacks. Here Nodes 6 and 18 are set as DoS attackers, and Nodes 3 and 19 are set as deception attackers.

\begin{figure}[!h]
	\centering
	\includegraphics[width=1\linewidth]{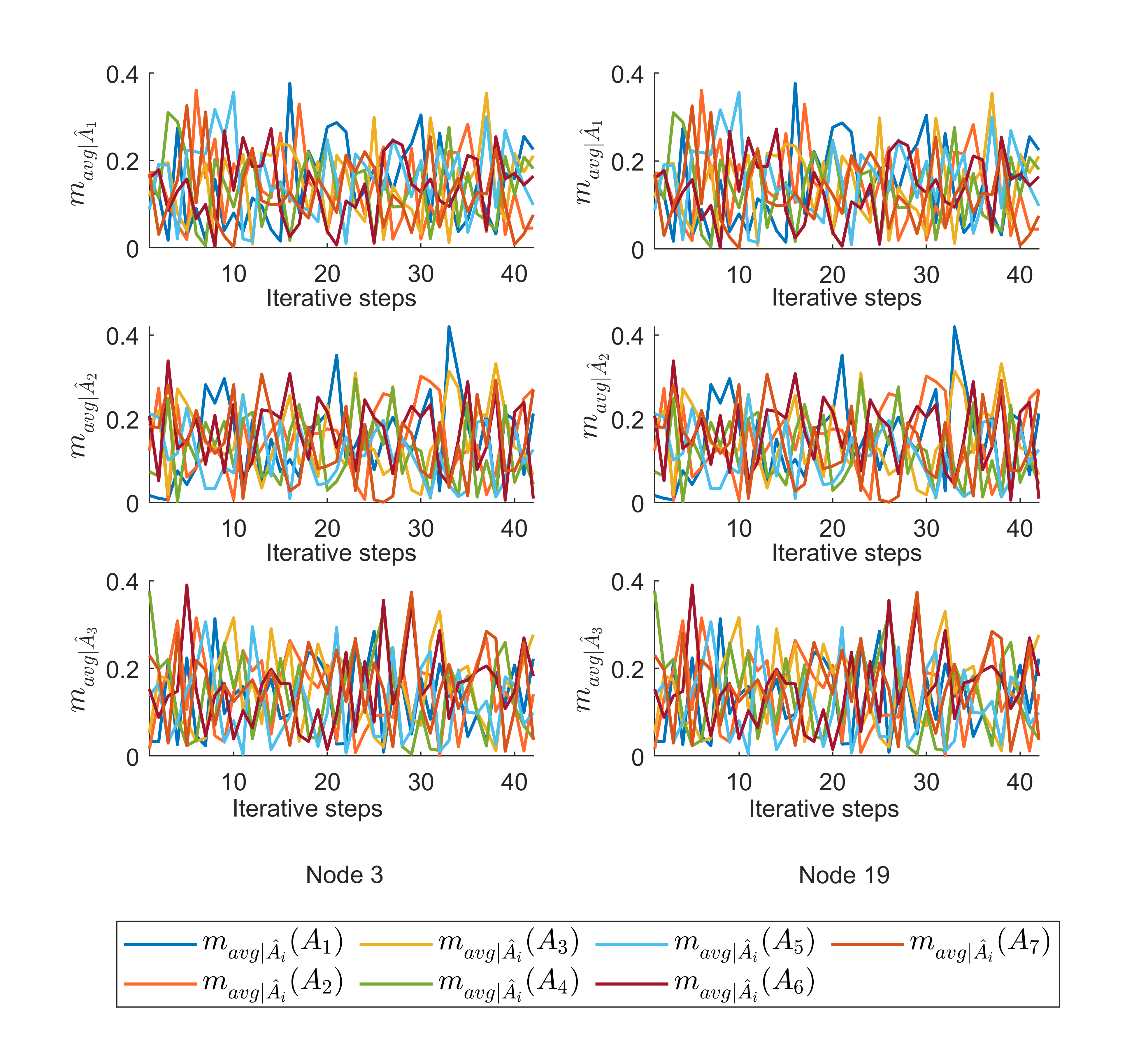}
	\caption{The WAVCCME sent by deception attackers during the consensus.}
	\label{fig:AttackCEF_DeceptionData}
\end{figure}

To simulate the effect of electromagnetic interference on the platform sensor measurements, this test sets Nodes 1 to 16 to provide normal evidence and Nodes 17 to 20 to provide abnormal evidence. The deception attackers, Nodes 3 and 19, then tamper with the sent data. The data they send at different moments are shown in Fig.\ref{fig:AttackCEF_DeceptionData}.

\begin{figure*}[!h]
	\centering
	\includegraphics[width=1\linewidth]{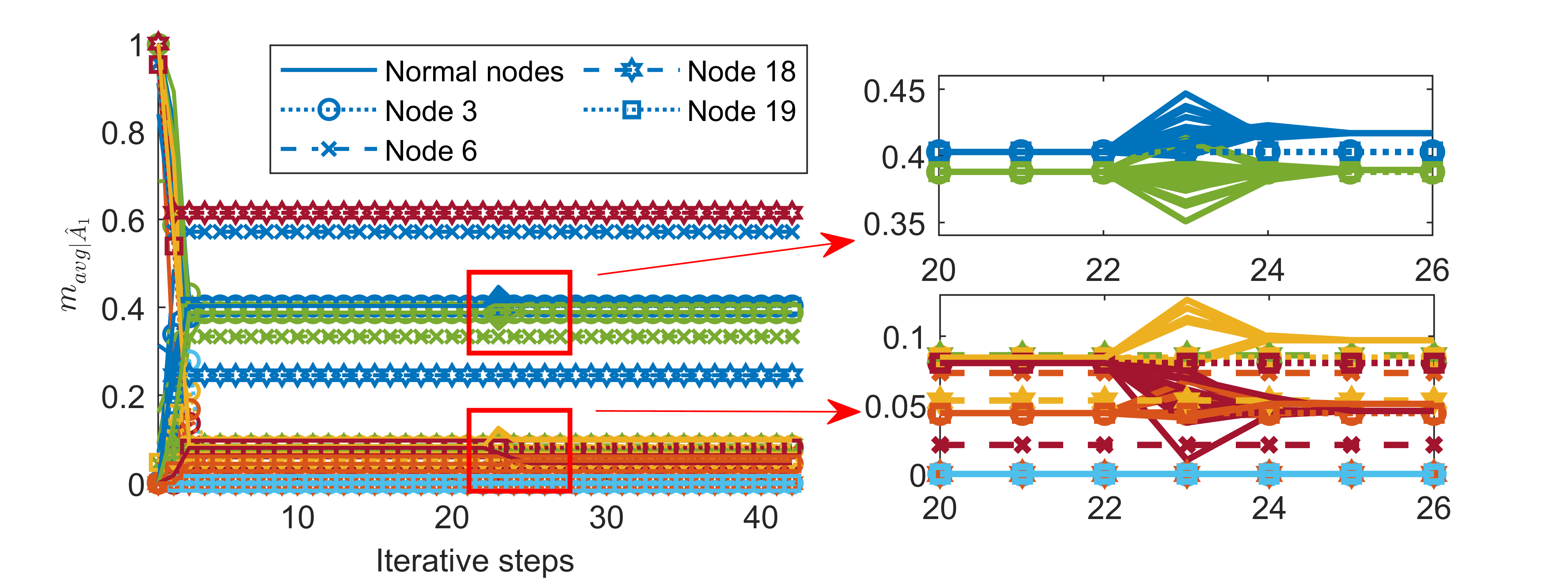}
	\caption{The WAVCCME curves for each node when the condition is event $\hat{A}_1$.}
	\label{fig:AttackCEF_NodeStateConvergenceCurve1}
\end{figure*}
\begin{figure*}[!h]
	\centering
	\includegraphics[width=1\linewidth]{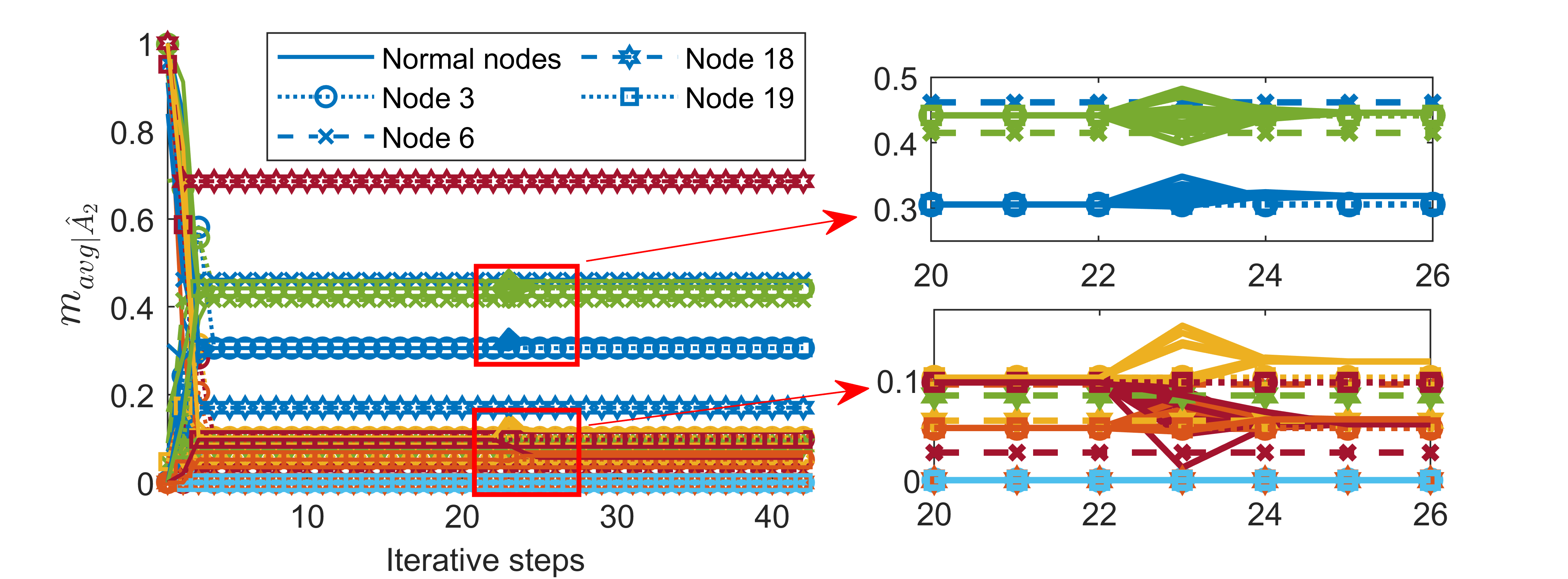}
	\caption{The WAVCCME curves for each node when the condition is event $\hat{A}_2$.}
	\label{fig:AttackCEF_NodeStateConvergenceCurve2}
\end{figure*}
\begin{figure*}[!h]
	\centering
	\includegraphics[width=1\linewidth]{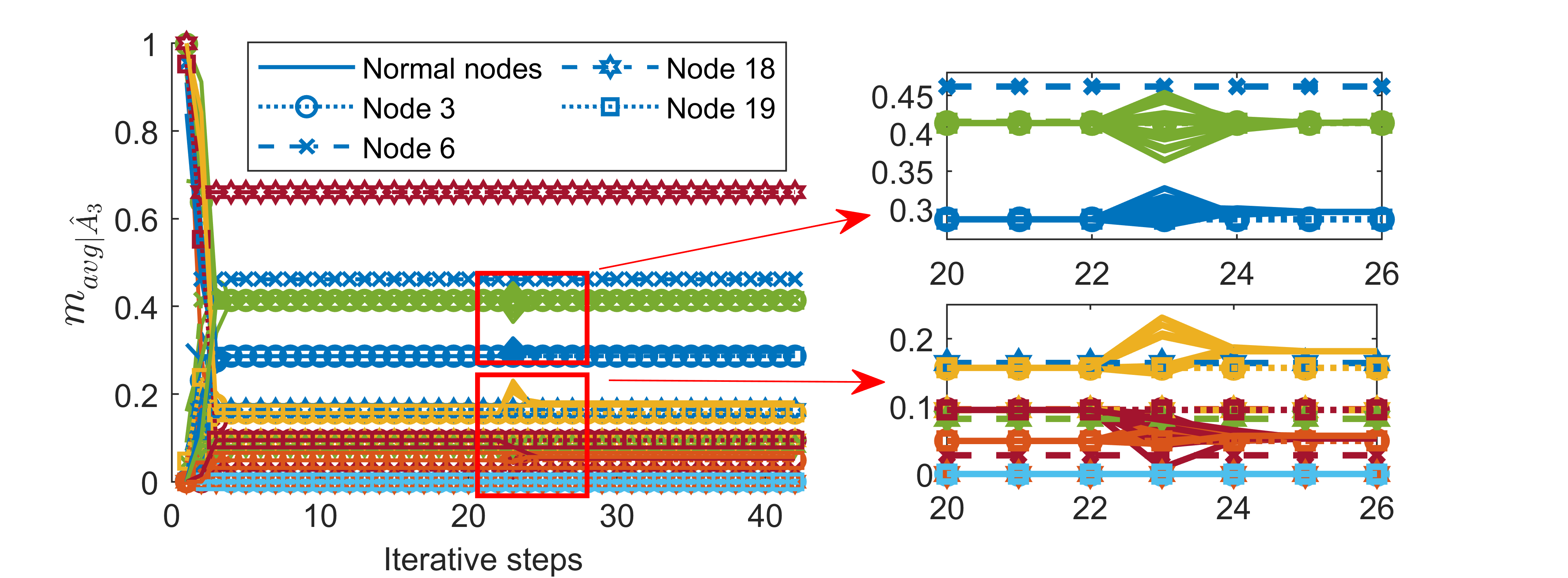}
	\caption{The WAVCCME curves for each node when the condition is event $\hat{A}_3$.}
	\label{fig:AttackCEF_NodeStateConvergenceCurve3}
\end{figure*}
The WAVCCME is a matrix of $(2^{|\Omega|}-1)\times |\Omega|$. In this test, its columns are the weighted sum of multi-source evidence corresponding to the conditions $\hat{A}_1$, $\hat{A}_2$, and $\hat{A}_3$, respectively. Fig.\ref{fig:AttackCEF_NodeStateConvergenceCurve1}-Fig.\ref{fig:AttackCEF_NodeStateConvergenceCurve3} show the WAVCCME for 20 nodes under different conditions. It is observed that all the normal nodes successfully converge to a consistent WAVCCME. The deception attackers are able to receive data from other nodes but are unable to eliminate and compensate for the data errors caused by the attack, and hence they end up in the same state but different from the normal nodes. The DoS attackers are unable to receive and transmit any valid information, so they converge to state values that are very different from the rest of the nodes.

\begin{table}[!h]
	\caption{Fusion results of different nodes.}
	\label{tab:Attack_FusionRessultAgainstAttack}
	\centering
	\begin{tabular}{cccc}
		\toprule [1.25pt]
		& $\{\hat{A}_1\}$ & $\{\hat{A}_2\}$ &$\{\hat{A}_1,\hat{A}_2\}$\\
		\midrule [0.5pt]
		Normal Nodes & 0.9261 & 0.0688 & 0.0051 \\
		Node 3                  & 0.8859 & 0.1105 & 0.0035 \\
		Node 6                  & 0.8482 & 0.0369 & 0.1149 \\
		Node 18                & 0.0000 & 0.4428 & 0.5367 \\
		Node 19                & 0.8859 & 0.1105 & 0.0035 \\
		Centralized                     & 0.9261 & 0.0688 & 0.0051 \\
		\bottomrule [1.25pt]
	\end{tabular}
\end{table}
This test uses the fusion result obtained by running Eqs.(\ref{eq:ICEF_Summary_1})-(\ref{eq:ICEF_Summary_4}) in a centralized manner as the reference benchmark for CEFAC. In this process, the identities of the attacking nodes are assumed to be known, and their evidences are not fused. Table \ref{tab:Attack_FusionRessultAgainstAttack} lists the fusion results of different nodes. It is noticed that the fusion result obtained by normal nodes is the same as the reference benchmark, which indicates that the proposed method effectively excludes the influence of network attacker evidence on distributed evidence fusion.
\section{Conclusion}
\label{sec: conclusion}
This paper studies the distributed fusion problem of multi-source evidence under cyber-adversarial behaviors, such as eavesdropping and attacks, and proposes a credible evidence fusion algorithm against cyberattacks to guarantee the credibility of fusion at three levels: evidence privacy protection, attack identification and compensation, and evidence high-conflict cancellation. By transforming the distributed fusion of evidence into average consensus of WAVCCME, this paper greatly simplifies the process of distributed evidence fusion and reduces the computational complexity. In order to adapt to the fusion demand under the directed graph, this paper designs a weight-encrypted state decomposition and reconstruction strategy to protect the privacy of node evidence: decompose node states into multiple random sub-states and send them to different neighbors to defend against internal eavesdroppers and encrypt the sub-states' weights in the reconstruction process to guard against eavesdroppers outside the system. In order to avoid attackers from influencing the fusion result, the proposed update rule with state correction helps to realize the consensus of normal nodes' WAVCCME. Simulation results show that the proposed algorithm provides a significant improvement over existing methods in terms of fusion accuracy and operational efficiency and accurately excludes attacker evidence from fusion.

%

\section*{Acknowledgements}
This work is supported by the National Natural Science Foundation of China under Grant 61873205.


\bibliography{reference}
\end{document}